\documentclass[rmp, aps]{revtex4}

\usepackage{natbib}
\usepackage{verbatim}
\usepackage{latexsym}

\usepackage{amsmath}
\usepackage{natbib}

 \usepackage{amssymb,amsfonts,amsmath}

\usepackage[pdftex]{graphicx}

\RequirePackage[normalem]{ulem} 
\RequirePackage{color}\definecolor{RED}{rgb}{1,0,0}\definecolor{BLUE}{rgb}{0,0,1} 

\def\J{J}
\def\hrho{\hat{\rho}}
\def\bx{{\bf x}}
\def\by{{\bf y}}
\def\bz{{\bf z}}
\def\bu{{\bf e}}
\def\x{x}

\def\z{z}
\def\bR{{\bf R}}
\def\H{H}
\def\M{M}
\def\q{q}
\def\c{c}
 
 \def\Q{Q}
 
\def\K{K}
\def\C{C}

\def\S{S}
\def\BS{B_S}
\def\qs{\q_\S}
\def\cs{\c_\S}

\def\E{E}
\def\LL{\ell_<}
\def\LG{\ell_>}
\def\LT{\tilde{\ell}}
\def\FG{F_>}
\def\FL{F_<}
\def\TG{\tau_>}
\def\TL{\tau_<}
\def\TT{\tilde{\tau}}
\def\PF{P_F}
\def\PS{P_S}
\def\DT{\tilde{D}}

 \def\pr{{\cal P}}
 
\def\be{\begin{equation}}
\def\ee{\end{equation}}

\def\lb{\left}
\def\rb{\right}

\begin{document}

\title{The acceleration of evolutionary spread by long-range dispersal}

\author{ Oskar Hallatschek} \affiliation{Biophysics and Evolutionary
  Dynamics Group, Department of Physics, University of California,
  Berkeley, CA 94720} \author{Daniel S. Fisher}
\affiliation{Departments of Applied Physics, Biology, and
  Bioengineering, Stanford University, Stanford, CA 94305}

\begin{abstract}
  The spreading of evolutionary novelties across populations is the
  central element of adaptation.  Unless population are well-mixed
  (like bacteria in a shaken test tube), the spreading dynamics not
  only depends on fitness differences but also on the dispersal
  behavior of the species. Spreading at a constant speed is generally
  predicted when dispersal is sufficiently short-ranged. However, the
  case of long-range dispersal is unresolved: While it is clear that
  even rare long-range jumps can lead to a drastic speedup, it has
  been difficult to quantify the ensuing stochastic growth
  process. Yet such knowledge is indispensable to reveal general laws
  for the spread of modern human epidemics, which is greatly
  accelerated by aviation.  We present a simple iterative scaling
  approximation supported by simulations and rigorous bounds that
  accurately predicts evolutionary spread for broad distributions of
  long distance dispersal.  In contrast to the exponential laws
  predicted by deterministic 'mean-field' approximations, we show that the
  asymptotic growth is either according to a power-law or a stretched
  exponential, depending on the tails of the dispersal kernel. More
  importantly, we provide a full time-dependent description of the
  convergence to the asymptotic behavior which can be anomalously slow
  and is needed even for long times.  Our results also apply to
  spreading dynamics on networks with a spectrum of long-range links
  under certain conditions on the probabilities of long distance
  travel and are thus relevant for the spread of
  epidemics. 
 \end{abstract}

\maketitle


Humans have developed convenient transport mechanisms for nearly any
spatial scale relevant to the globe. We walk to the grocery store,
bike to school, drive between cities or take an airplane to cross
continents. Such efficient transport across many scales has changed
the way we and organisms traveling with us are distributed across the
globe~\cite{ruiz2000global,suarez2001patterns,brockmann2006scaling,gonzalez2008understanding,rhee2011levy}. This
has severe consequences for the spread of epidemics: Nowadays, human infectious
diseases rarely remain confined to small spatial regions, but instead
spread rapidly across countries and continents by travel of infected
individuals~\cite{brockmann2013hidden}.

Besides hitchhiking with humans, small living things such as microbes
or algae are easily caught by wind or sea currents, resulting in
passive transport over large spatial
scales~\cite{brown2002aerial,mccallum2003rates,d2004mixing,martin2003phytoplankton,perlekar2010population}. Effective
long distance dispersal is also wide-spread in the animal kingdom,
occurring when individuals primarily disperse locally but occasionally
move over long distances. And such animals, too, can transport smaller
organisms.


These active and passive mechanisms of long-range dispersal are
generally expected to accelerate the growth of fitter mutants in
spatially extended populations. But how can one estimate the resulting
speed-up, and the associated spatio-temporal patterns of growth?  When
dispersal is only short-range, the competition between mutants and
non-mutated (``wild-type'') individuals is local, confined to small
regions in which they are both present at the same time. As a
consequence, a compact mutant population emerges that spreads at a
constant speed, as first predicted by Fisher and Kolmogrov and
colleagues~\cite{fisher1937wave,KPP37}: such selective sweeps are slow
and dispersal limited. In the extreme opposite limit in which the
dispersal is so rapid that it does not limit the growth of the mutant
population, the competition is global and the behavior the same as for
a fully-mixed (panmictic) population: mutant numbers grow
exponentially fast.  It is relevant for our purposes to note that in
both the short-range and extreme long-range cases, the dynamics after
the establishment of the initial mutant population is essentially
deterministic.


When there is a broad spectrum of distances over which dispersal
occurs, the behavior is far more subtle than either of the
well-studied limits. When a mutant individual undergoes a long
distance dispersal event -- a jump -- from the primary mutant
population into a pristine population lacking the beneficial mutation,
this mutant can found a new satellite  sub-population, which can
then expand and be the source of further jumps, as shown in Fig. 1B
and 1C. Thus, long-range jumps can dramatically increase the rate of
growth of the mutant population.  Potentially, even very rare
long-range exceptionally long jumps could be important. If this is the
case, then the stochastic nature of the jumps that drive the dynamics
will be essential.

While evolutionary spread with long-range jumps has been simulated
stochastically in a number of biological
contexts~\cite{mollison1972rate,
  clark2003estimating,filipe2004effects,cannas2006long,marco2011comparing,brockmann2013hidden},
few analytic results have been obtained on the ensuing stochastic
dynamics~\cite{mollison1972rate,biskup2004scaling,chatterjee2013multiple}.
Most analyses have resorted to deterministic
approximations~\cite{mollison1977spatial,kot1996dispersal,neubert2000demography,mancinelli2002superfast,del2003front,brockmann2007front,del2009truncation},
which are successful for describing both the local and global
dispersal limits. Yet, in between these extreme limits, stochasticity
drastically changes the spreading dynamics of the mutant
population. This is particularly striking when the probability of
jumps decays as a power-law of the distance.  Just such a distance
spectrum of dispersal is characteristic of various biological
systems~\cite{levandowsky1997random,klafter1990microzooplankton,atkinson2002scale,fritz2003scale,ramos2004levy,dai2007short}.
We will show that the behavior is controlled by a balance between the
rarity and the potential effectiveness of long distance jumps and the
whole spectrum of jump distances can matter. The goal of this paper is
to develop the theory of stochastic spreading dynamics when the
dispersal is neither short range nor global.

Long distance dispersal can occur either on a fixed network, or more
homogeneously in space.  For simplicity, we focus on the completely
homogeneous case, and then show that many of the results also apply
for an inhomogeneous transportation network with hubs between which
the long distance jumps occur. For definiteness, we consider for most
of the paper the evolutionary scenario of the spread of a single
beneficial mutation, but, by analogy, the results can be applied to
other contexts, such as the spread of infectious disease or of
invasive species.


\begin{figure*}
\includegraphics[width=.95\textwidth]{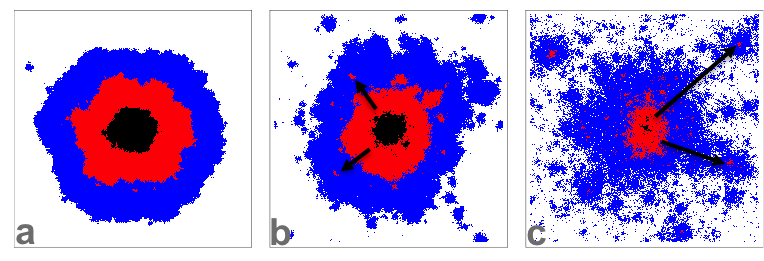}
\caption{Evolutionary spread sensitively depends on the dispersal
  behavior of individuals. For a broad class of models with
  'short-range migration', the mutant subpopulation expands at a
  constant speed that characterizes the advance of the
  mutant-nonmutant boundary.  With long-range dispersal the spread is
  much faster. The figure depicts two-dimensional simulations for the
  simple case of a jump distribution that has a broad tail
  characterized by a power law exponent $-(\mu+2)$. The figures show
  the distribution of the mutant population at the time when half of
  the habitat is occupied by mutants. The color of a site indicates
  whether a site was filled in the first (black), second (red) or last (black)
  third of the total run time. A) When the jump distribution decays
  sufficiently rapidly, $\mu>3$, the asymptotic growth resembles, in
  two dimensions, a disk growing at a $\mu$--dependent constant
  speed. B) For $2<\mu<3$, satellite seeds become clearly visible and
  these drive super-linear power law growth. These seeds were
  generated by long-range jumps, as indicated by arrows. C) The
  dynamics is changed drastically for $0<\mu<2$, becoming controlled
  by very long distance jumps, which seed new expanding satellite
  clusters. As a result of this ``metastatic'' growth, the spreading
  is faster than any power law, though markedly slower than
  exponential. The figures were created by the simulations described
  in the main text, with parameters $\mu=3.5$ in A), $\mu=2.5$ in B)
  and $\mu=1.5$ in C). \label{fig:illustrate-sweep-dynamics}}
\end{figure*}

\subsection*{Basic model}
\label{sec:minimal-model}

The underlying model of spatial spread of a beneficial mutant is a
population in a $d$-dimensional space with local competition that keeps the
population density constant at $\hrho$ and 
with a  probability that any individual jumps to any particular
point a distance $r$ away of $\J(r)$ per time per area, per length or
per volume.  At a very low rate, mutants can appear that
have a selective advantage, $s$, over the original population.  A lattice
version of this model is more convenient for simulations (and for
aspects of the analysis): each lattice site represents a ``deme" with
fixed population size, $\hat{n}\gg \frac{1}{s}$ with the competition
only within a deme and the jump migration between demes.
Initially, a single mutant occurs and if, as occurs with probability
proportional to $s$, it survives stochastic drift to establish, it
will take over the local population. When the total rate of migration
between demes is much slower than this local sweep time, the spatial
spread is essentially from demes that are all mutants to demes that
are all the original type.

Short jumps result in a mutant population that spreads spatially at a
roughly constant rate.  But with long-range jumps, new mutant
populations are occasionally seeded far away from the place where they
came, and these also grow. The consequences of such long jumps is the
key issue that we need to understand.  As we shall see, the
interesting behaviors occur when the jump rate has a power-law tail at
long distances, specifically, with $\J(r)\sim 1/r^{d+\mu}$ (with
positive $\mu$ needed for the total jump rate to be finite).  Crudely,
the behavior can be divided into two types: linear growth of the
radius of the region that the mutants have taken over, and faster than
linear growth.  In Fig.~\ref{fig:illustrate-sweep-dynamics}, these two
behaviors are illustrated via simulations on two dimensional
lattices. In addition to the mutant-occupied region at several times,
shown are some of the longest jumps that occur and the clusters of
occupied regions that grow from these.  In
Fig.~\ref{fig:illustrate-sweep-dynamics}A, there are no jumps that are
of comparable length to the size of the mutant region at the time at
which they occur, and the rate of growth of the characteristic linear
size $\ell(t)$ of the mutant region --- loosely its radius --- is
roughly constant in time, i.e. $\ell(t)\sim t$. In
Fig.~\ref{fig:illustrate-sweep-dynamics}B and C, $J(r)$ is longer
range and very long jumps are observed.  These result in faster-than
linear growth of the radius of the mutant region, as shown.  Before
developing analytic predictions for the patterns of evolutionary
spread, we report our simulation results in detail.

\section{Results}
\label{sec:results}
\subsection{Simulated spreading dynamics}
\label{sec:snapshots}
We have carried out extensive simulations of a simple lattice model
that can be simulated efficiently. The sites form either a
one-dimensional, of length $L$, or two-dimensional $L\times L$ square
array with periodic boundary conditions. As it is the spreading
dynamics at long times that we are interested in, we assume that the
local sweeps in a deme are fast compared to migration. We can then
ignore the logistic growth process within demes, so that when jumps
occur and establish a new mutant population it is saturated in the new
deme by the next time step.  Therefore, it is convenient to lump
together the probability of an individual to jump, the density of the
population from which the jumps occur, and the probability
(proportional to $s$) that the mutant establishes a new population: we
define $G(r)\equiv s\hrho \J(r)$ so that $d^dr d^dr' G(|\bx-\by|)$ is
the rate at which a saturated mutant population near $\bx$ nucleates a
mutant population near $\by$. In each computational timestep, we pick
a source and target site randomly such that their distance $r$ is
sampled from the (discretized) jump distribution $G(r)$ -- with the
$d^dr$ a lattice site. If the source site is a mutant and the target
site a wild type, the identity of the target site is updated to
mutant. We measure time in units of $L^d$ timesteps. See SI Sec.~S1 for
more details on the simulation algorithm.

\begin{figure}
  \includegraphics[width=\columnwidth]{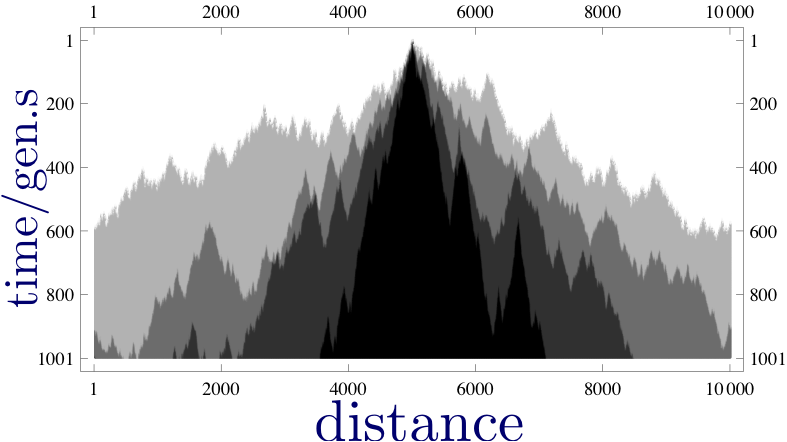}
  \includegraphics[width=\columnwidth]{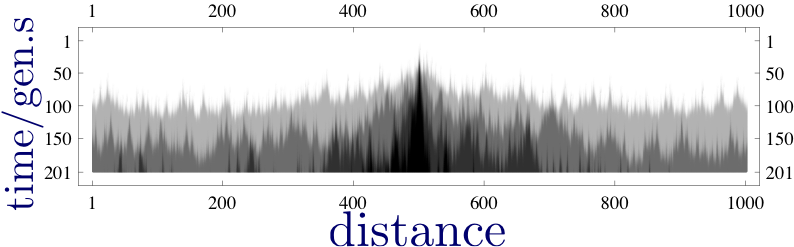}
  \caption{Stochastic growth of a mutant population over time in
    one dimension. Each level of grey represents a single simulation
    run. A) Power law regime: The values of  $\mu$  are
    $1.5, 1.6, 1.7, 2.0$ in order of increasing darkness. B) Regime of very fast growth with $\mu$
    near 1. The values of $\mu$  are $0.8, 0.9, 1.0, 1.1$ in order of
    increasing darkness. \label{fig:snapshots}}
\end{figure}

\begin{figure}
  \center{\includegraphics[width=.9\columnwidth]{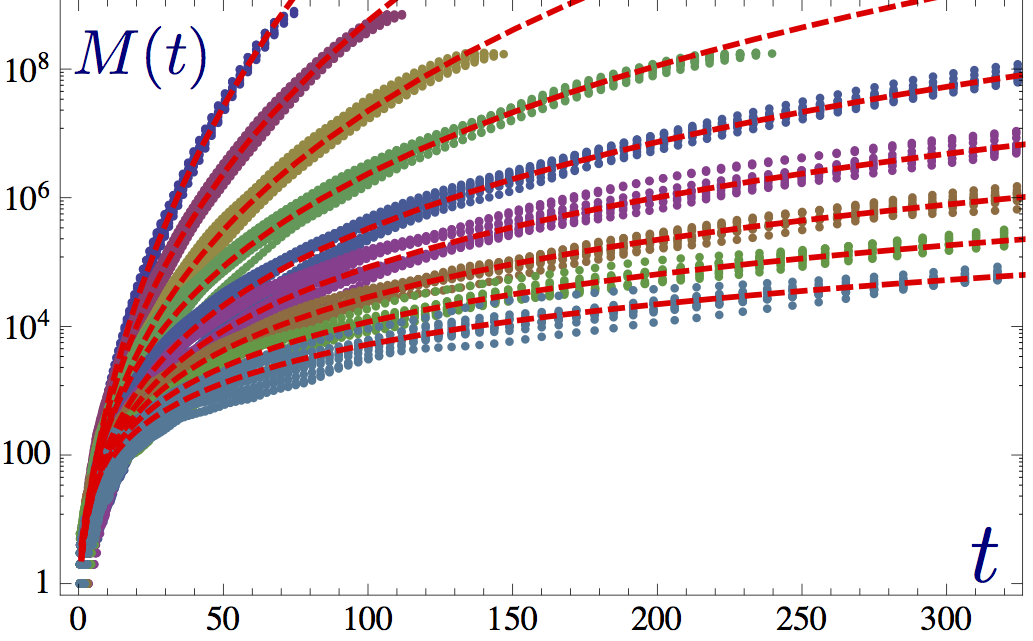}}\\
  \center{\includegraphics[width=.9\columnwidth]{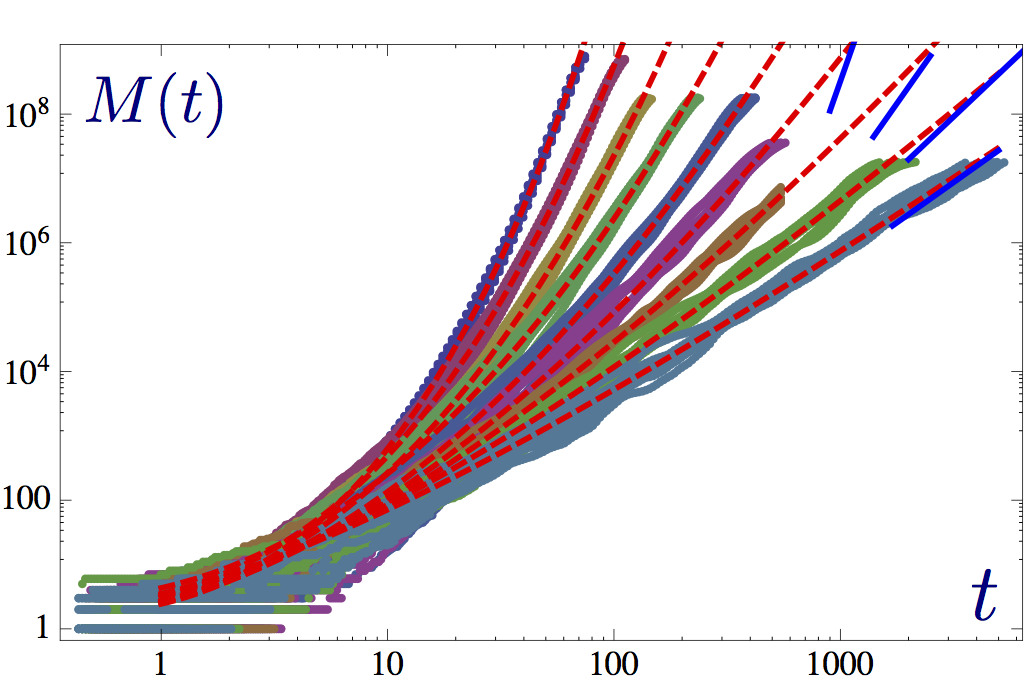}}
  \caption{Summary of the quantitative spreading dynamics in one spatial dimension.
    The total number, $M(t)$, of mutant sites is plotted as a function
    of time $t$, for various long-range jump kernels. Each colored
    cloud represents data obtained from $10$ runs for a given jump
    kernel. The data  are for $\mu\in\{0.6, 0.7, 0.8, 0.9, 1.0,
    1.1, 1.2, 1.3, 1.4\}$ from top to bottom. Red dashed lines
    represent predictions  obtained from equation
    \eqref{eq:solution-recurrence-relation} with fitted magnitude
    scales for $M$ and $t$. In the double logarithmic plot at the bottom, 
    short blue lines indicate the predicted asymptotic power-law behavior for
    $\mu>1$. For $\mu=1.1$ and $\mu=1.2$, the dynamics is still far
    away from the asymptotic power law, which is indicative of the very
    slow crossover. \label{fig:v1-spreading}}
\end{figure}

The growth of mutant populations generated by our simulations is best
visualized in a space-time portrait. Fig.~\ref{fig:snapshots}A and
Fig.~\ref{fig:snapshots}B show the overlaid space-time plots of
multiple runs in the regimes $1.5<\mu<2$ and $0.8<\mu<1.1$,
respectively. Fig.~\ref{fig:v1-spreading} shows the growth dynamics of
the mutant population over large time and length scales for various
values of $\mu$. For $\mu \gtrsim 1.4$, the dynamics clearly
approaches a power law. For $\mu \lesssim 0.7$, the simulations are
consistent with stretched exponentials. The intermediate regime $0.7
<\mu < 1.3$ is elusive, as one cannot extract a clear asymptotic
behavior on the time scales feasible in simulations. The behavior in
two-dimensions is qualitatively similar, as shown in the SI
Fig.~S1.

To explain these dynamics in detail, we develop an analytical
theory that is able to predict not only the asymptotic growth dynamics
but also the crucial transients.

\subsection{Breakdown of Deterministic Approximation}
\label{sec:determ-appr}
Traditionally, analyses  of spreading dynamics start with a deterministic approximation of
the selective and dispersal dynamics --- ignoring both stochasticity and the discreteness of individuals. To setup consideration of the
actual stochastic dynamics, we first give results in this
deterministic approximation and show that these exhibit hints of why
they break down. When the jump rate decreases exponentially or faster
with distance, the spread is {\it qualitatively} similar to simple
diffusive dispersal and the extent of the mutant population expands
linearly in time.  However when the scale of the exponential fall-off is long, the speed, $v$, is faster than the classic
result for local dispersal, $v=2\sqrt{Ds}$, which only depends on the
diffusion coefficient, $D=\frac{1}{2d}\int r^2 J(r) d^dr$:
Instead, there is a characteristic range of jump distances determined
by $s$ and $\J(r)$
that dominate the spread. As the dispersal range gets longer for fixed $D$, the
dominant jumps occur at lower rates suggesting that the stochasticity
may become more important.

If $\J(r)$ has a long  tail, in particular, $\J(r)\sim 1/r^{d+\mu}$, the
spread in the deterministic approximation becomes exponentially fast
with the radius of the region taken over by the mutant population,
$\ell(t) \sim \exp(dst/(d+\mu))$.  This is almost as fast as in a fully
mixed population, with the growth rate of the total mutant population
size slower only by a factor of $\frac{d}{d+\mu}$ which approaches
unity as $\mu\to 0$, the point at which the spatial structure becomes
irrelevant. The origin of this exponential growth for power-law
$\J(r)$ is the deterministic feeding of the populations far away by jumps from near the
origin: this immediately produces a finite population density at all
distances, $r$.   
After a time of order $1/s$ has passed, the exponential growth of the
local population takes off, that is, further jumps to that region no
longer matter.  As in this time $1/s$ the expected total number of
jumps to the {\it whole} region further than $r$ from the origin is
only of order $1/(sr^\mu)$, the probability that {\it any} jumps have
occurred is very small for large $r$ and the deterministic
approximation must fail~\cite{mollison1972rate}.

With local dispersal, the deterministic approximation is a good
starting point with only modest corrections to the expansion speed at
high population density, the most significant effect of
stochasticity being fluctuations in the speed of the
front~\cite{van2003front}.  At the opposite extreme of jump rate
independent of distance, the deterministic approximation is also good
with the mutant population growing as $e^{st}$ and fluctuations only
causing stochastic variability and a systematic reduction in the
coefficient which arise from early times when the population is small.
Surprisingly, in the regimes intermediate between these two, the
deterministic approximation is not even qualitatively reasonable.

\subsection{Iterative scaling argument}
\label{sec:self-cons-theory}
We assume that, at long times, most of the sites are filled out to some
distance scale $\ell(t)$ and that the density decreases sufficiently steeply for
larger distances, such that the total mutant population, $M(t)$,  is
proportional to $\ell^d(t)$. The validity of the this assumption
follows from more accurate analyses given in the Appendix.  We call
the crossover scale $\ell(t)$ the {\it core
  radius} or ``size"  of the mutant population.

In the dynamical regimes of interest, the core population grows
primarily because it ``absorbs'' satellite clusters, which themselves
were seeded by jumps from the core population.  We now show that the
rate of seeding of new mutant satellite clusters and the growth of the core
populations by mergers with previously seeded clusters has to satisfy
an iterative condition that enables us to determine the typical
spreading dynamics of the mutant population.

\begin{figure}
  \includegraphics[width=.95\columnwidth]{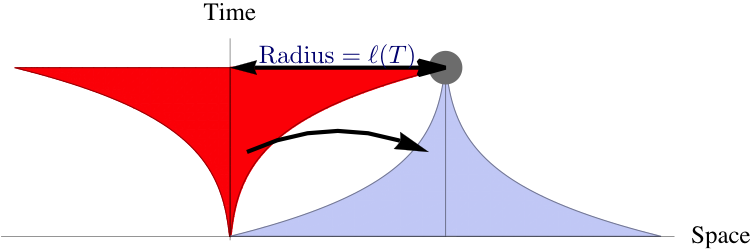}
  \caption{Sketch of the  growth of the compact core of
    a cluster (red) due to long-range jumps. For the (gray) point at
    distance $\ell(t)$ to be occupied at time $t$, a seed typically
    must become established somewhere in the blue space-time region
    (``target funnel'') by means of a long-range jump (black arrow)
    from the red ``source" region.  This schematic leads to the
    iterative scaling approximation in Eq.~\eqref{eq:1}. Note that the
    concavity of the source-funnel geometry, leading to a gap between
    red and blue regions, is key to our arguments and enables
    neglecting effects of jumps into the gap
    region.  \label{fig:funnel-antifunnel}}
\end{figure}


It is convenient to illustrate our argument using a space-time
diagram, Fig.~\ref{fig:funnel-antifunnel}, in which the growth of the
core has the shape of a funnel.  Now consider the edge of this funnel
at time $T$ (gray circle in the figure). The only way that this edge
can become populated is by becoming part of a population sub-cluster
seeded by an appropriate long-range jump at an earlier time. To this
end, the seed of this sub-cluster must have been established somewhere
in the inverted blue funnel in Fig.~\ref{fig:funnel-antifunnel}. This
``target'' funnel has the same shape as the space-time portrait of the
growing total population, but its stem is placed at $(\ell(T),T)$ and
the mouth opens backward in time. Notice that if $\ell(t)$ grows
faster than linearly, space-time plots of the growing cluster and the
funnel have concave boundaries: this necessitates a jump from the
source to the funnel of length much longer than $\ell(T/2)$, as shown.

In $d$ dimensions, the consistency of growth and seeding thus requires
\begin{equation}
  \label{eq:1}
  \int_0^{T}dt \int_{\mathcal{B}_{\ell(t)}}d^d\bx
  \int_{\mathcal{B}_{\ell(T-t)}}d^d\by\; G\left(|\ell(T) \bu+\by-\bx| \right) \approx 1 \;,
\end{equation}
where $\mathcal{B}_\ell$ denotes a $d$ dimensional ball of radius
$\ell$ centered at the origin, and we have taken the point of interest to be $\bR=\bu\ell(T)$ with $\bu$ a unit vector in an arbitrary direction. The kernel $G(r)$ represents the rate per $d-$dimensional volume
of (established) jumps of size $r$. 

\subsection{Asymptotic results for power law jumps}
\label{sec:power-law-regime} 
We now show that the asymptotic growth dynamics is essentially
constrained by the above iterative scaling argument. Specifically, although
the argument is more general, we consider a power law jump
distribution
\begin{equation}
  \label{eq:2}
G(r) \approx \frac{\epsilon}{r^{d+\mu}} \ \ \ {\rm with}  \ \ \  \mu<d+1 
\end{equation}
for large enough $r$. 

For the {\it intermediate-range case} $d<\mu<d+1$, equations \eqref{eq:1} and \eqref{eq:2} exhibit the
asymptotic scaling solution
\begin{equation}
  \label{eq:plaws}
  \ell(t)\sim A_\mu (\epsilon t)^\beta \ \ \ {\rm with}  \ \ \  \beta =\frac{1}{\mu-d}\ >1 \;,
\end{equation}
the {\it form} of which could have been guessed by dimensional
analysis.  Inserting the ansatz \eqref{eq:plaws} into equation
\eqref{eq:1} determines the pre-factor $A_\mu$ in this iterative
scaling approximation, up to an order-unity coefficient: see details
in the Appendix. Interestingly, the value $A_\mu$ depends very
sensitively on $\mu$ and runs from 0 to $\infty$ as $\mu$ passes
through the interval from $d$ to $d+1$. This can be seen in in SI
Fig.~S2, where $A_\mu$ is plotted as a function of $\mu$ for $d=1$: it
drops very steeply, as $A_\mu\sim 2^{-2/(\mu-1)^2}$, for $\mu\searrow
1$, and diverges as $A_\mu\sim 1/(2-\mu)$ for $\mu\nearrow 2$. As we
discuss below, these singularities are a manifestation of intermediate
asymptotic regimes that lead to very slow convergence to the
asymptotic behavior.

We now turn to the (very) {\it long-range case}, $0<\mu<d$, for which a direct
solution to \eqref{eq:1} cannot be found (and the dimensional analysis argument gives nonsense). However, much can be learned by approximating \eqref{eq:1}
using $G[|\ell(T) \bu+\by-\bx|]\approx G[\ell(T)]$, anticipating the very rapid growth and thus likely smallness of $x$ and $y$ compared to $\ell(T)$:
\begin{equation}
  \label{eq:approx-self-consist-condition}
   G[\ell(T)] \int_0^{T}dt \ell^d(t)\ell^d(T-t) \approx 1 \;,
\end{equation}
(ignoring two factors 
from the angular integrations).  With $\ell(t)$ growing
sub-exponentially, the largest contributions will come from $t\approx
\frac{1}{2}T$. Once we have found the form of $\ell(t)$, the validity
of the ansatz can be tested by checking whether the $\ell(T)$ is much
larger than $\ell(T/2)$. Indeed, for $\mu<d$ the solution to
\eqref{eq:approx-self-consist-condition} is a rapidly growing
stretched exponential, \be
\label{eq:explaws} 
\ell(t)\sim \exp(B_\mu t^\eta) \ \ \ {\rm with} \ \ \ \eta =\frac{
\log[2d/(d+\mu)]}{\log 2}\ , 
\ee 
which can be checked by direct
insertion into equation \eqref{eq:approx-self-consist-condition}.
Notice that as $\mu\searrow 0$, $\eta\nearrow 1$ and $\ell(t)$ grows
exponentially as for a flat distribution of long-range jumps that
extends out to the size of the system: i.e. the globally-mixed limit. In the opposite limit of
$d-\mu$ small, $\eta \sim d-\mu$ and the coefficient $B_\mu$ diverges as shown below.
We  note that the asymptotic
stretched-exponential growth for $\mu<d$ also arises in models of ``chemical distance" and certain types of spatial spread for network models with a similar power-law distribution of long distance connections: however the pre-factors in the
exponent are different~\cite{biskup2004scaling,biskup2004graph}. We discuss the connections between these in Sec.~\ref{sec:heterogen}.

For the {\it marginal case}, $\mu=d$, the  asymptotic
behavior is similarly found to be: 
\begin{eqnarray}
  \label{eq:marginal}
  \ell(t)\sim \exp\left[\log^2(t)/(4d\log(2))\right] \;,
\end{eqnarray}
We now show that this behavior also represents an important
\emph{intermediate} asymptotic regime which dominates the dynamics
over a wide range of times for $\mu$ close to $d$: this is the source
of the singular behavior of the coefficients $A_\mu$ and $B_\mu$ in
this regime.

Our source-funnel argument obviously neglects jumps that originate
from the not-fully-filled regions outside the core radius
$\ell(t)$. An improved version of the funnel argument is presented in
the SI Sec.~S2~C, which also allows us to estimate the probability of
occupancy outside the core region. Further, we present in the Appendix
outlines of rigorous proofs of lower and upper bounds for the
asymptotic growth laws in one dimension, including the slow crossovers
near the marginal case.  The linear growth for $\mu>3$ in one
dimension has been proven by Mollison~\cite{mollison1972rate}.  After
the present paper was essentially complete, we became aware of a
recent preprint by Chatterjee and Dey~\cite{chatterjee2013multiple}
who obtained rigorous bounds in all dimensions for the leading
asymptotic behaviors in the three super-linear regimes.  Our bounds
are somewhat tighter than theirs, including the coefficient and
leading corrections to the asymptotic behavior in the marginal case,
the absence of logarithmic pre factors in the power-law regime, and
the full crossovers for $\mu$ near $d$ in one dimension: comparisons
are discussed in the Appendix.

\subsection{Crossovers and Beyond Asymptopia}
\label{sec:transients}
Asymptotic laws are of limited value without some understanding of
their regime of validity, especially if the approach to the asymptotic
behavior is slow.  And such knowledge is crucially needed to interpret
and make use of results from simulations.

We first consider the short time behavior when the total rate of long
jumps is small: i.e., when $\epsilon$, the coefficient of $G(r)\approx
\epsilon/r^{d+\mu}$ is small.  Short jumps result in diffusive motion
and linear growth $\ell(t) \approx v_0t$ with $v_0$ determined by the
details of the selective and diffusive dynamics.  Long jumps start to
become important after enough time has elapsed that there have been at
least some jumps of lengths of order $\ell(t)$: i.e. when $\epsilon
t\ell(t)^d /\ell(t)^{\mu}\gg1$: this occurs after a cross-over time
$t_\times\sim [v_0^{\mu-d}/\epsilon]^{1/(d+1-\mu)}$ at which point
$\ell\sim \ell_\times \sim [v_0/\epsilon]^{1/(d+1-\mu)}$.  At longer
times and distances, one can measure lengths and times in units of
these crossover scales, defining $\lambda\equiv\ell/\ell_\times$ and
time $\theta=t/t_\times$ and expect that the behavior in these units
will not depend on the underlying parameters.  [Note that this
separation in short-time linear growth and long-time regimes can also
be done for more general $G(r)$ although then the behavior will depend
on the whole function --- the crossover on distances of order
$\ell_\times$, and the super-linear behavior on the longer distance
form.]

At times much longer than $t_\times$, there is a slow crossover
close to the
boundary between the stretched exponential and power law regimes. Thus we must take a closer look at the dynamics in the vicinity of the marginal
case, $\mu=d$. 
As $\mu\searrow d$, 
the integrand in equation
\eqref{eq:approx-self-consist-condition} develops a sharp peak at $t=T/2$:
half-way between the bounds of the integral.  Laplace's
method can then be used to approximate the integral leading to a 
simplified recurrence relation: 
\begin{equation}
  \label{eq:self-consistency-close-to-marginal-nondim}
  \lambda^{d+\mu}(\theta)\sim \theta \lambda\left(\theta /2\right)^{2d} \;,
\end{equation}
This is really only good to a numerical factor of order unity which
can be eliminated by rescaling $\theta$, and to a larger logarithmic
factor associated with the narrowness of the range of integration and
its dependence on $\theta$ and $\mu-d$: this we analyze in SI
Sec.~S2~A. Both of these are negligible if we focus on the behavior on
logarithmic scales in space and time --- natural given their
relationships.  Defining $\varphi\equiv \log_2(\lambda)$ and $z\equiv
\log_2(\theta)$, and taking the binary logarithm, $\log_2$, of
equation \eqref{eq:self-consistency-close-to-marginal-nondim}, yields
a linear recurrence relation that can be solved exactly.  Rescaling
$\lambda$ can be used to make $\varphi(0)=0$ whence
\begin{equation}
  \label{eq:solution-recurrence-relation}
  \frac{\delta^2}{2d}\varphi(z) =\frac{\delta z}{2d}+\left(1+\frac{\delta}{2d}\right)^{- z}-1 \;,
\end{equation}
where we introduced the variable $\delta=\mu-d$, which measures the
distance to the marginal case.  

The asymptotic scaling for $\delta>0$ reproduces the earlier predicted
power law regime, \eqref{eq:plaws}, and yields the pre-factor $\log
A_\mu\approx - 2 d \log(2)/\delta^2$, up to correction that is
subdominant for small $\delta$ (c.f.  SI
Fig.~S3). For $\delta<0$, the asymptotics yields the
stretched exponential in \eqref{eq:explaws}, and fixes the pre-factor
$B_\mu\approx 2\log(2)\delta^{-2}$, which could not be obtained from
the basic asymptotic analysis carried out above.

The singular pre-factors for $\delta\to0$ give warnings of breakdown
of the asymptotic results except at very long times. This peculiar
behavior is the consequence of an intermediate asymptotic regime that
dominates the dynamics close to the marginal case. This leads to slow
convergence to the eventual asymptotic behavior for $\mu$ near $d$.
The asymptotic scaling can be observed only on times and length such
that
\begin{equation}
  \label{eq:asymptotic-limit}
  \log_2(\theta)\gg 2d/|\delta| \ \ \ {\rm and} \ \ \ \log_2(\lambda)\gg 2d/\delta^2\;.
\end{equation}
On smaller times, the dynamics is similar to the marginal case,
\eqref{eq:marginal}. The rapid divergence of the {\it logarithm} of
the time after which the asymptotic results obtain, make it nearly
impossible to clearly observe the asymptotic limits: in
one-dimensional simulations this problem occurs when $|\delta|<0.3$,
as us clearly visible in Fig.~\ref{fig:v1-spreading} and it is likely
even harder to observe in natural systems. This underscores the need for
the much fuller analysis of the spreading dynamics as via
\eqref{eq:solution-recurrence-relation}.

\begin{figure}
  \includegraphics[width=.9\columnwidth]{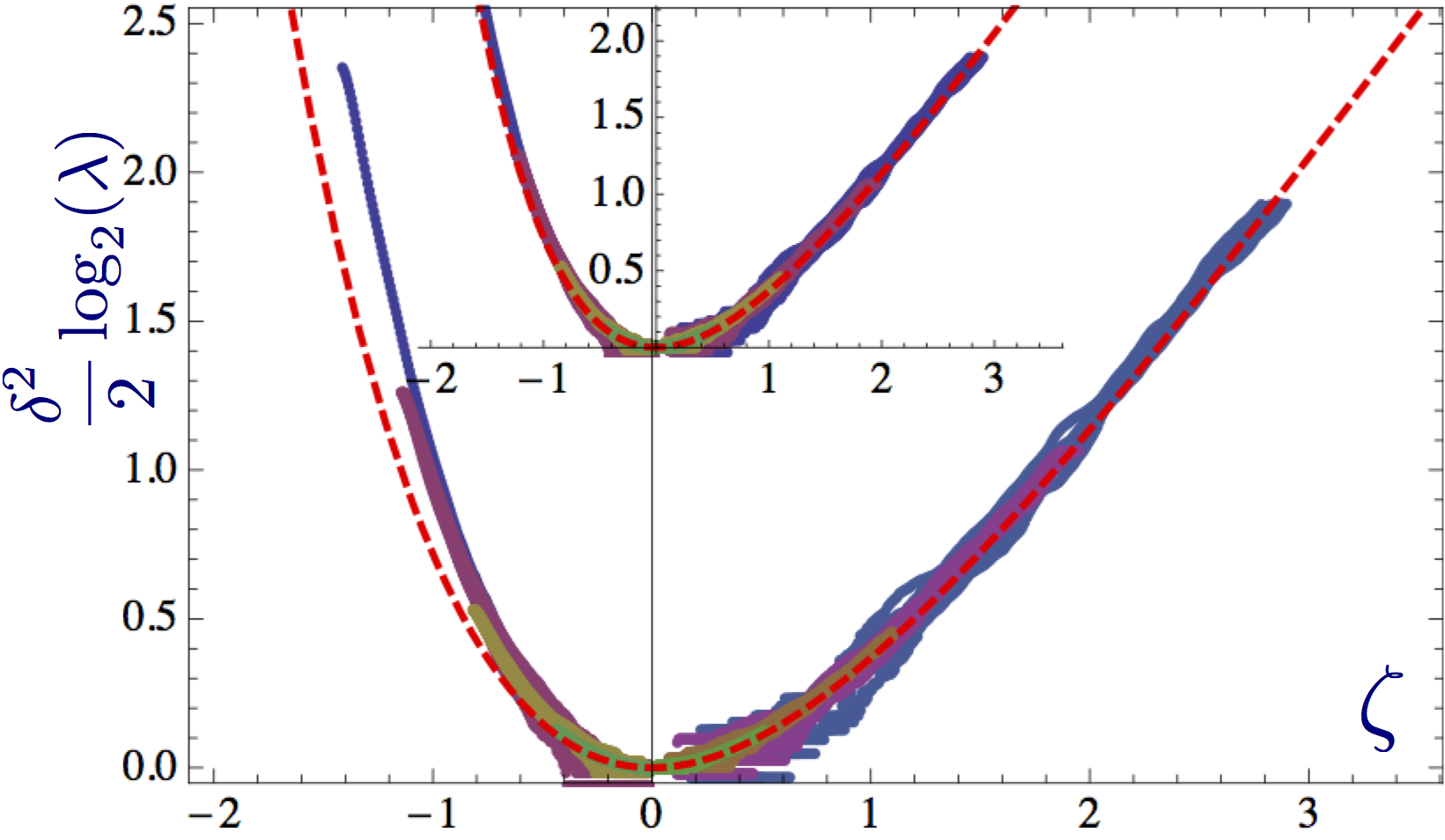}
  \caption{Data for different $\mu=d+\delta$ are predicted to collapse
    on a scaling plot close to the marginal case $\mu=d$ between the
    stretched-exponential and power-law growth regimes, here demonstrated
    for the one-dimensional case ($d=1$).  The main plot shows the
    rescaled sizes $\delta^2 \log_2(\lambda)/2$ of the mutant
    population versus rescaled log-time
    $\zeta\equiv\delta\log_2(\theta)/2$. The differently colored data
    sets correspond to $10$ realizations with power law exponents
    $\mu\in\{0.6, 0.7, 0.8, 0.9, 1.1, 1.2, 1.3, 1.4\}$ (same colors as
    in Fig.~\ref{fig:v1-spreading}a). The simulated data collapse
    reasonably well with the dashed red line, representing the
    predicted scaling function $\chi(\zeta)=\exp(-\zeta)+\zeta-1$. The
    inset depicts the same data scaled slightly differently away from the scaling regime such that
    the horizontal axis for $\delta<0$ shows $\eta \log(\theta)$ (with
    $\eta$ defined in \eqref{eq:explaws}), upon which the data
    collapse improves. Note that $\eta \log(\theta)\approx  -\zeta$ as
    $\delta\to0$ so that the stretched exponential form is recovered for $\zeta$ large and negative. \label{fig:scaling-plot-a}}
\end{figure}

While the dynamics at moderate times will be dominated by the initial
growth characteristic of the marginal case, we expect
\eqref{eq:solution-recurrence-relation} to be a good description of
the universal dynamics at large $z=\log_2 \theta$ even when $\delta$ is
small.  The limit $z\to\infty$ while $\zeta\equiv \delta z/2d$ fixed
is particularly interesting, as the solution
\eqref{eq:solution-recurrence-relation} then reduces to a scaling form
\begin{equation}
  \label{eq:scform}
  \frac{\delta^2\varphi}{2d}\approx \chi\left(\frac{\delta
      z}{2d}\right)
\end{equation}
with 
\be
\chi(\zeta)=\exp(-\xi)+\xi-1\ . 
\ee
This scaling form allows us to test by simulations our analytic
results across all intermediate asymptotic regimes, by plotting data
obtained for different $\delta$ in one scaling plot, see
Fig.~\ref{fig:scaling-plot-a}.  To make the approximation uniformly
valid in both the scaling regime and at asymptotically long times
outside of it, we can simply replace, for $\delta<0$, the scaling
variable by $\zeta\equiv -\log(\theta)\eta$ (with $\eta$ defined in
\eqref{eq:explaws}): the scaling form \eqref{eq:scform} will then be
valid up to corrections that are small compared to the ones given in
all regimes: We thus use this form for the scaling fits in the inset
of Fig.~\ref{fig:scaling-plot-a}, plotting data obtained for different
$\delta$ in one scaling plot, thereby testing our solution across all
intermediate asymptotic regimes.

\subsection{Heterogeneities and Dynamics on Networks}
\label{sec:heterogen}

Thus far we have considered spatially uniform systems, in which the
jump probability between two points only depends on their separation.
However, long distance transport processes may be very
heterogeneous. An extreme example is airplane travel which occurs on a
network of links between airports with mixtures of short and long
distance flights, plus local transportation to and from airports.  A
simple model is to consider each site to have a number of connections
from it, with the probability of a connection between each pair of
sites a distance $r$ away being $\C(r)$, independently for each pair:
note that although the network is heterogeneous, statistically, the
system is still homogeneous as the connection probability does not
depend on position.  If the rate at which jumps occur across a
connection of length $r$ is $\H(r)$, then, averaged over all pairs of
sites, the rate of jumps of distance $r$ is $\J(r)=\C(r)\H(r)$.  How
similar is this to a homogeneous model with the same $\J(r)$, in
particular if $\J(r)\sim 1/r^{d+\mu}$? If there are a large number of
possible connections along which the key jumps can occur to get from a
source region of size $\sim\ell(T/2)$ to a funnel of similar size a
distance $\ell(T)$ away, then the fact that these only occur to and
from a small fraction of the sites should not matter for the large
length scale behavior.  The number of such connections is $n_c\sim
\ell(T/2)^{2d}\C(\ell(T))$ with, making the ansatz that, as in the the
homogeneous case, $T\ell(T/2)^{2d}\J(\ell(T))\sim 1$ (ignoring
subdominant factors) one has $n_c \sim 1/[T\H(\ell(T))]$.  Thus the
condition for our results to be valid asymptotically is that $\H(r)
\ll 1/ \tau(r)$ with $\tau(r)$ the inverse of the function $\ell(t)$:
i.e., in the exponential, marginal, and power-law cases, respectively,
that $\H(r) \ll \log(r)^{-1/\eta}$, $\H(r) \ll \exp(-\sqrt{4d\log
  2\log r})$, and $\H(r)\ll 1/r^{\mu-d}$.

If there are insufficient number of connections for the heterogeneity
of the network to be effectively averaged over, the behavior changes.
The extreme situation is when there is a distance-independent rate for
jumps along the longest connection out of a site: i.e. $\H(r) \to
const.$. In this case, jumps along the path with the shortest number
of steps, $S$, to get from the origin to a point $R$ will reach that
point in a time proportional to $S$ i.e. $\tau(R)\sim S(R)$.  The
geometrical problem of obtaining the statistics of $S(R)$ has been
analyzed by Biskup~\cite{biskup2004scaling,biskup2004graph,DaSilveira}.  For
$\mu>d$, $ S\sim R$ and long jumps do not matter, while for $\mu<d$,
$S\sim (\log R)^{1/\eta}$ with the same exponent $\eta$ as in the
homogeneous case we have analyzed. The difference between this result
and ours is only in the power-law-of-$T$ pre-factor of $\ell(T)$
arising from the integral over time: this does not exist in the
extreme network limit.  In the marginal case, $\mu=d$, $T\sim S \
R^\alpha$ with $\alpha$ dependent on the coefficient of the power-law
decay of the connection probability. 

 If the probability of a jump along a long distance connection decays
 with distance but more slowly than $1/\tau(r)$, the behavior is
 similar: for $\mu<d$ again the ubiquitous stretched exponential
 behavior occurs, while for $\mu>d$ there are too few connections only
 if $\C(r) < 1/r^{2d}$ in which case the number of steps and the time
 are both proportional to the distance.  The marginal cases we have
 not analyzed further. 

 For natural transport processes, the probabilities of long dispersal
 events will depend on both the source and the destination.  If the
 heterogeneities are weak on large length scales, our results still
 obtain.  But if there are sufficiently strong large scale
 heterogeneities, either in a spatial continuum or in the network
 structure (i.e. location of nodes and links and the jump rates along
 these, or hub-spoke structure with multiple links from a small subset
 of sites), then the spatial spread will be heterogeneous even on
 large scales: how this reflects the underlying heterogeneities of the
 dispersal has to be analyzed on a case by case basis.

\section{Discussion}
\label{sec:discussion}
We have studied the impact of long-range jumps on evolutionary
spreading using the example of mutants that carry a favorable genetic
variant. To this end, we analyzed a simple model in which long-range
jumps lead to the continual seeding of new clusters of mutants, which
themselves grow and  send out more migrant mutants. The ultimate merging
of these satellite clusters limits the overall growth of the mutant
population, and it is a balance of seeding and merging of sub-clusters
that controls the spreading behavior.

To classify the phenomena emerging from this model, we focussed on
jump distributions that exhibit a power law tail. We found that,
with power law jumps, four generic behaviors are possible on
sufficiently long times: The effective radius of the mutant population
grows either at constant speed, as a super-linear power law of time,
as a stretched exponential, or simply exponentially depending on
the exponent, $d+\mu$, of the power law decay of the jump probability.  These predictions are in contrast to
deterministic approximations that predict exponential growth for power-law decaying jump kernels
\cite{mollison1977spatial,kot1996dispersal,mancinelli2002superfast,del2003front}.
In dimensions more than one, the results also contradict the naive expectation from dynamics of neutral dispersal, that a finite diffusion coefficient is sufficient for conventional behavior (in this context, finite speed of spreading): specifically,  for $d+1 > \mu >2$, $D<\infty$ but the spread is super-linear --- indeed stretched exponential for $d>\mu>2$.  That super-linear dynamics can occur for $\mu<2$ is not surprising as even a migrating individual undergoes a Levy flight: more surprising is that this occurs even when the dynamics of individuals is, on large scales, like a normal random walk.

The breakdowns of both deterministic  and diffusive expectations are
indicative of the importance of fluctuations: the dynamics is dominated by very rare --- but not too rare --- jumps: roughly, the most unlikely that occur  at all up to that time.  One of the consequences of this control by the rare jumps is the relatively minor role played by the selective fitness advantage, $s$, of the mutants.  With short-range dispersal, the speed is $v\approx \sqrt{Ds}$, and for global dispersal, $\log M \approx st$. 
By
contrast, in the regime of power law growth, the
asymptotic growth of the mutant population becomes (to leading order)
independent of the speed of the growth of individual clusters although when
individual clusters grow more slowly, the asymptotic regime is reached
at a later time. 
In the stretched exponential regime,
the growth of sub-clusters sets the cross-over time from linear to stretched
exponential, and thus determines the pre-factor in the power law that
characterizes the logarithm of the mutant population size.

\begin{figure}
  \includegraphics[width=.9\columnwidth]{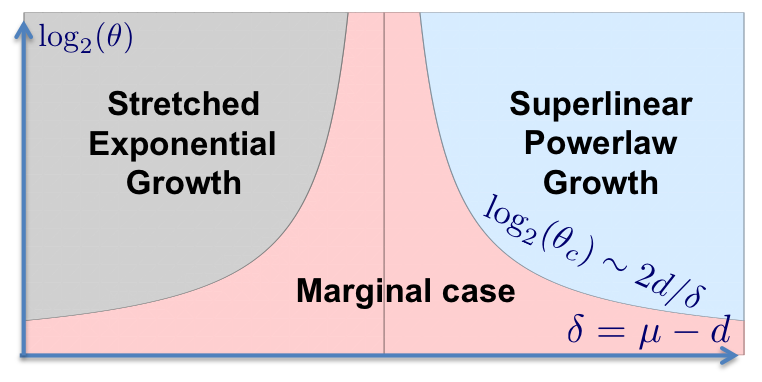}
  \caption{Close to the marginal case, $\mu=d$, the spreading dynamics
    exhibits three behaviors. On asymptotically large times,
    either stretched exponential growth for $-d<\delta\equiv
    \mu-d<0$ or superlinear power law growth for $0<\delta<1$ occurs. However,
    the approach to the asymptotic regime is extremely slow for $\mu$ close to  $d$.
    For log-times $\log_2(\theta)\ll 2d/|\delta|$,
    the behavior is controlled by the dynamics of the marginal
    case. Note that cross-over behavior also obtains near to the
    borderline between linear and super-linear behavior at
    $\mu=d+1$, cf. SI Sec.~S2~B. 
  }
  \label{Fig:phase-diagram}
\end{figure}

An important feature of our results is that the approach to the
asymptotic laws is very slow in the vicinity of the marginal cases, as
illustrated in Fig.~\ref{Fig:phase-diagram}.  Consider for instance
the two-dimensional case, where we have asymptotically stretched
exponential growth for $\mu<2$: For $\mu=1.8$ ($\mu=1.6$), the
epidemics has to run for times $\theta\gg 10^6$ $(10^3)$ to reach the
asymptotic regime. By that time, however, the mutant population with
$\lambda\gg 10^{30}$ $(10^7)$ would have certainly spread over the
surface of the earth. To describe the full cross-over dynamics, we
developed an approximate recurrence relation based on the geometric
source-funnel argument described in the illustration
Fig.~\ref{fig:funnel-antifunnel}. This predicts a universal cross-over
function for the transient dynamics near $\mu=d$ that can be uncovered
by simulations for different power law exponents collapsed onto one
scaling plot, c.f. Fig.~\ref{fig:scaling-plot-a}.  The good agreement
of the predicted scaling function and simulation results provides
strong support for the iterative scaling approximation.  The rigorous
bounds whose proofs are outlined in the SI Sec.~S3, provides further
support.  Understanding this crossover is essential for making sense
of, and extrapolating from, simulations as asymptotic behavior is not
visible until enormous system sizes even when the exponent $\mu$ is
more than $0.2$ from its marginal value.

Another benefit of the ability of the simple iterative scaling
argument to capture well non-asymptotic behavior, is that it can
be used in cases in which the dispersal spectrum of jumps is not a
simple power-law, e.g.  with a crossover from one form to another as a
function of distance.  And the heuristic picture that it gives rise to
-- an exponential hierarchy of time scales separated by roughly
factors of two --- is suggestive even in more complicated situations.
That such a structure should emerge without a hierarchal structure of
the underlying space or dynamics is perhaps surprising.

\subsection{Potential applications and dynamics of epidemics}

 Our primary  aim biologically  is the qualitative and semi-quantitative 
understanding that emerges from consideration of the simple models and analyses of these, especially demonstrating how rapid spatial spread of beneficial mutations or other biological novelty can be even with very limited long-range dispersal. 
As the models do not depend on any detailed information about the biology or dispersal mechanisms, they can be considered as a basic null model for spreading dynamics in {\it physical}, rather than more abstract network, space.  

The empirical literature suggests that power-law spectra of spatial dispersal are
wide-spread in the biological
world~\cite{levandowsky1997random,klafter1990microzooplankton,atkinson2002scale,fritz2003scale,ramos2004levy,dai2007short}.
Although these are surely neither  a  constant power-law over a wide
range of scales, nor spatially homogeneous, our detailed results are
not directly applicable. But, as discussed above,  our iterative
scaling argument is more general and can be applied with more complicated distance dependence, anisotropy, etc.  Furthermore, some of the heterogeneities of the dispersal  will be averaged out for the overall spread, while affecting when mutants are likely to arrive at particular locations.   

For dispersal via hitchhiking on human transport, either of pathogens
or of commensals such as fruit flies with food, the apparent
heterogeneities are large because of the nature of transportation
networks, although data suggest that dispersal of humans can be reasonably approximated by power-law jumps~\cite{brockmann2006scaling}. 
Whether or not transport via a network with hubs at many scales fundamentally changes the dynamics of an expanding population of hitchhikers from that with more homogeneous jump processes, depends on the nature of how  the population expands.  

For spread of a human epidemic, there are several possible scenarios.
If the human population is reasonably uniform spatially, and the
chances that a person travels from, say, their home to another
person's home is primarily a function of the distance between these
rather than the specific locations, then whether or not the properties
of the transportation network matter depends on features of the
disease.  If individuals are infectious for the whole time the
outbreak lasts and if transmission is primarily at end points of
journeys rather than enroute --- for example HIV --- then the
transportation network plays no role except to provide the spatial
jumps.  At the other extreme but still within an SI model, is if
individuals living near hubs are more likely to travel (or even if
destinations near hubs are more likely), and, more so, if infections
are likely to occur enroute, in which case the structure of the
transportation network --- as well as of spectra of city sizes, etc
--- matters a great deal.  In between these limits the network (or
lack of it in places) may matter for initial local spread but at
longer times the network structure may effectively average out and the
dynamics be more like the homogeneous models.  The two opposite limits
and behavior in between these, together with the specific network
model we analyzed with jumping probabilities depending on distance
even in the presence of a connection --- as is true from airports ---
all illustrate an important point: geometrical properties of networks
alone rarely determine their properties: quantitative aspects, such as
probabilities of moving along links and what exists at the nodes, are
crucial.



More complicated epidemic models can be discussed within the same
framework: The model discussed thus far corresponds to an SI model,
the most basic epidemic model, which consists of susceptible and
infected individuals only. Many important epidemics are characterized
by rather short infectious periods, so that one has to take into
account the transition from infected to recovered: SIR models. This
changes fundamentally the geometry of our space time analysis,
illustrated in Fig.~\ref{fig:funnel-antifunnel}. While the target
funnel remains a full funnel, the source funnel becomes hollow: The
center of the population consists mostly of fully recovered
individuals, whose long-range jumps are irrelevant. The relevant
source population of infected individuals is primarily near the
boundaries of the funnel. This leads to a break in the time-symmetry
of the argument. As a result, the spreading crosses over from the
behavior described above to genuine SIR behavior. The SIR dynamics is
closely related to the scaling of graph-distance in networks with
power-law distributions of link
lengths~\cite{biskup2004scaling,biskup2004graph,DaSilveira}, as
recently shown by one- and two-dimensional
simulations~\cite{grassberger2013sir,grassberger2013two}.  In
particular, the limited time of infectiousness causes wave-like
spreading at a constant speed for $\mu>d$. But, importantly, the
spreading velocity is controlled by the SI dynamics we have studied
until a time of order the infectious period.  Analogous crossovers
from SI to SIR occur also with longer range jumps.  More generally,
other complications can be discussed in our framework, and we expect
new behaviors depending on how they modify the geometry of the source
to target-funnel picture.

\begin{figure*}
  \includegraphics[width=.95\textwidth]{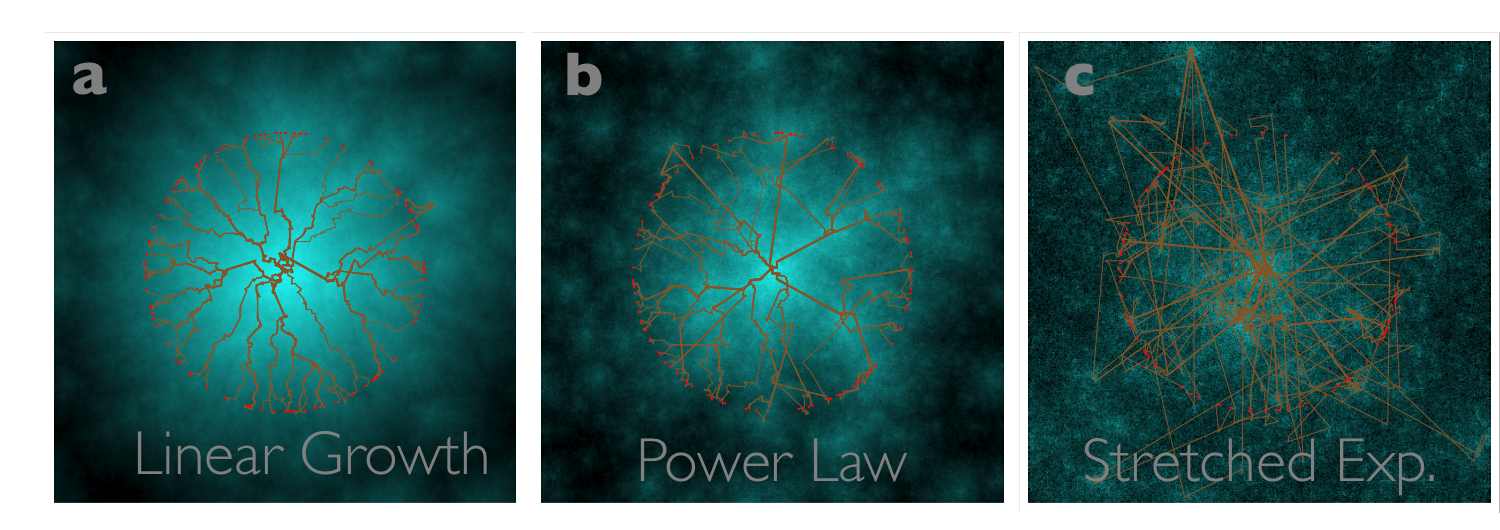}
  \caption{These figures depict coalescent trees, or infection trees
    in epidemiology, generated by power-law dispersal. The
    subfigures a, b and c each represent one simulation run with
    parameter $\mu=3.5$, $\mu=2.5$ and $\mu=1.5$, respectively. For
    each run, 100 lattice sites indicated as red points
    equidistantly from the start point at the origin were sampled. For each labeled site the path that led to its colonization is plotted. The resulting coalescence trees are
    characteristic of the three different regimes, and reveal the
    long-range jumps that drove the colonization process. The
    background  represents the colonization process  in
    time with the color of each site indicating its colonization time
    (light blue=early, dark=late). \label{fig:infection_trees}}
\end{figure*}

  As a last note, our analyses naturally provide information on the
  typical structure of infection (or coalescent) trees.  For a given
  site, the path of jumps by which the site was colonized can be
  plotted as in Fig.~\ref{fig:infection_trees}.  Doing this for many
  sites yields coalescent (or infection) trees, which can reveal the
  key long-range jumps shared by many lineages. For actual epidemics,
  combining spatio-temporal sampling of rapidly evolving pathogens
  with whole genome sequencing is now making it possible to construct
  such infection trees.  For inference purposes, it would therefore be
  interesting to analyze more about the statistical properties of such
  infection trees and how they depend on the dispersal properties,
  network structure, and other features of epidemic models.


\begin{acknowledgments}
This work was partially supported by the DFG via HA 5163/2-1, NSF via DMS-1120699
and PHY-1305433.
\end{acknowledgments}


\newpage
\appendix
\renewcommand{\thepage}{S\arabic{page}}  
\renewcommand{\thesection}{S\arabic{section}}   
\renewcommand{\theequation}{S\arabic{equation}}   
\renewcommand{\thetable}{S\arabic{table}}   
\renewcommand{\thefigure}{S\arabic{figure}}

\setcounter{equation}{0}
\setcounter{page}{1}

\setcounter{figure}{0}

\section{Simulation details}
\label{sec:simulation-details}

\subsection{Simulation Algorithm}
\label{sec:simulation-algorithm}

The state of the population is described by a linear array of $N$
sites with periodic boundary conditions. $N$ is chosen large enough so
that end-effects can be ignored (typically between $10^7$ and $10^8$
sites).  Each site has the identity of either mutant or
wildtype. Initially, the whole population is wildtype except for the
central site, which is occupied by mutants.

In each computational timestep, a source site $A$ and target site $B$
are chosen randomly such that their distance $r$ is sampled from a
probability density function with a tail $\mu r^{-(1+\mu)}$ at large
$r$ (see below). If $A$ is mutant and $B$ a wild type then $B$ turns into mutants
and seeds a new mutant cluster. We use the convention that $N$
time-steps --- i.e. an average of one jump attempt per site --- comprise one unit of time, or effective ``generation".  The rate of long-range jumps
should be thought of as representing the product of the probability to
establish a new cluster per jump and the jump rate per generation per
site.

\subsection{Jump size distribution}
\label{sec:jump-size-distr}
In our simulations, the distance $X$ of a long-range jump was
generated as follows. First, draw a random number $Y$ within $(0,1)$,
and calculate the variable
\begin{equation}
  \label{eq:generatejumps1}
  X=\left[Y \left(L^{-\mu}-C^{-\mu}\right) +C^{-\mu} \right]^{-1/\mu}\;.
\end{equation}
where $C$ is a cutoff (see below) and $L$ is the system size. This
generates a continuous PDF
\begin{equation}
  \label{eq:generatejumps2}
  \Pr(X=x)=x^{-(\mu+1)} \frac{\mu (C L)^\mu}{L^\mu-C^\mu} 
\end{equation}
with $x$ values in $(C,L)$. The actual jump distance is obtained from
$X$ by rounding down to the next integer. (Note that, because the distribution
has a tail $\mu r^{-(1+\mu)}$, we have to choose $\epsilon=
\mu$ in equation (2) of the main text.)

For the one dimensional data in all graphs of this paper, we used $C=1$ and system
sizes ranging from $L=10^9$ for $\mu=0.6$ to $L=10^8$ for
$\mu=1.4$. For such large systems, the tail of the distribution is
well approximated by $p(x)\sim \mu x^{-(\mu+1)}$, as stated in
the main text. We also tested variations in the cutoff $C$. Using
$C=10$ or $C=100$ only affected the short-time dynamics and had very
little influence on the intermediate asymptotic or long distance behavior of the
system.

For our two-dimensional simulations, we draw jump sizes from the same
distribution as the one described above. We set the lower cutoff to
$C=1.5>\sqrt{2}$ to make sure that jumps reach out of the source
lattice point. After the jump size is drawn, the jump direction is chosen at
random. 

\section{Iterative scaling approximation: details and extensions}
\label{sec:Iterative-scaling-approx}

\subsection{Sub-dominant corrections from time integrals}
\label{sec:laplaces-method}

In analyzing the results from the iterative scaling approximation  we have ignored the effects of the range of $t$ around  $T/2$ that dominates the probabilities of occupation at time $T$.   This is valid for obtaining the leading behaviors of $\log \ell(t)$  in the large time limit, but there are corrections to these that can be larger than those that arise from the short-time small-length-scale  crossovers that we discussed in the text and do so further below.  For $\mu$ not much smaller than $d+1$, when the growth of $\ell$ is a modest power of time, the factor from the range of the time integral  is only of order unity and hence no worse than other factors -- including from the stochasticity --- that we have neglected.  But when $\ell(t)$ grows very rapidly, the range of $\frac{1}{2}T - t$ that dominates is much smaller than $T$ and the corrections are larger. 

With rapid growth of $\ell(t)$, a saddle point
approximation to the time integral is valid:  $\int_0^tdt\ell(t)\ell(T-t)\approx Tc_{T/2} 
\ell^2(T/2)$ with the pre-factor given by 
 \begin{equation}
  \label{eq:integral-laplace}
  c_t \approx \sqrt{\frac{2\pi}{-2t^2\partial_t^2\log \ell(t)}} 
\end{equation}
which, with the second derivative absorbing the $t^2$ factor, the
derivative part can be rewritten as $t^2\partial_t^2\log
\ell(t)=-\partial_{\log t}\log \ell +\partial^2_{\log t}\log \ell$.
With the asymptotic growth laws we have derived, this gives $c\sim
\sqrt{\mu-d}$ for $\mu-d$ small and positive, $c_t \sim 1/\sqrt{\log
  t}$ for $\mu=d$, and $c_t \sim t^{-\eta/2}$ for $\mu<d$.
Integrating up the effects of this over the scales yields the
following corrections in the various regimes: For $\mu>d$, the
coefficient, $A_\mu$, of $t^\beta$ is changed by a multiplicative
factor which is much less singular for $\mu \searrow d$ than that
already obtained. For $\mu=d$, \be \log [\ell(t)] \approx C[\log^2 t -
\log t \log\log t + {\cal O}(\log t)] \ee with $C=1/4d\log 2$, the
second term being new and the smaller correction term including the
effects of the small time
cross-over. 
For $0<\mu<d$, \be \log [\ell(t)] \approx B_\mu t^\eta -
\frac{1-\eta/2}{d-\mu}\log t \ee with the second term for $\mu
\nearrow d$ just that which occurs in the crossover regime analyzed in
Sec.~''Crossovers and Beyond Asymptopia'' (main text), where we showed that the coefficient $B_\mu$
diverges proportional to $1/(d-\mu)^2$.

\subsection{Pre-factors in power law regime}
\label{sec:power-law-regime-1}
In the main text, we mainly focussed on regimes  in which the
mutant growth is very much faster than linear, i.e. $\mu \lesssim
d+1/2$. This allowed us to approximate the integrals in the
iterative scaling approximation of equation (1) (main text) by the use of Laplace's
method. This saddle point approximation yields the correct scaling for
all exponents $\mu<d$, but (as we see below) incorrect pre-factors in
regimes where the actual growth is close to linear, i.e. in the power
law growth regime with $d+1>\mu\gtrsim d$.

To obtain a better estimate of the pre-factors in this power law
regime, it is helpful to directly solve for the asymptotics of the
iterative scaling argument in equation (1) (main text). Here, we
demonstrate how this can be done for $d=1$: Assume that most of the
weight in the integral comes from regions, where the jump kernel is
well approximated by its power-law tail described in equation (2)
(main Text). Given $\mu< 2$ (for $d=1$), this always holds at
sufficiently long times. Then, one has
\begin{equation}
  \label{eq:Oskar}
  \epsilon \int_0^{t}dt' H(t') \;,  
\end{equation}
where
\begin{eqnarray}
  \label{eq:Oskar2}
 \mu (\mu-1)H(t')&=& (\ell(t)-\ell(t')-\ell(t-t'))^{1-\mu} \nonumber \\
  & &-(\ell(t)-\ell(t')+\ell(t-t'))^{1-\mu} \nonumber \\
  & &+(\ell(t)+\ell(t')+\ell(t-t'))^{1-\mu} \nonumber\\
 & &-(\ell(t)+\ell(t')-\ell(t-t'))^{1-\mu} \;.
\end{eqnarray}
For $1<\mu<2$, equation \eqref{eq:Oskar} exhibits an asymptotic power
law solution
\begin{equation}
  \label{eq:plaws-glaw}
  \ell(t)=A_\mu (\epsilon t)^{1/(\mu-1)}\;.
\end{equation}
By inserting this ansatz into Eq.s~\eqref{eq:Oskar} and ~\eqref{eq:Oskar2},
we obtain the following result for the numerical pre-factor
\begin{equation}
  \label{eq:kappa-OH}
  A_\mu^{1/(\mu-1)}=\int_0^1dz \tilde R(z)
\end{equation}
with $\tilde R(z)$ being equal to Eq.~(\ref{eq:Oskar2}) with $\ell(t)$
replaced by $z^{1/(\mu-1)}$.

The resulting pre-factor is plotted as a function of $\mu-1$ in
Supplementary Fig.~\ref{fig:prefactors}. Notice that $A_\mu$ strongly
depends on the exponent $\mu$. It sharply drops for $\mu$ approaching
$1$, where it follows the asymptotics $2^{-2(\mu-1)^{-2}}$. On the
other hand, as $\mu$ approaches the other marginal case at $\mu=2$,
the pre-factor diverges as $A_\mu\sim (2-\mu)^{-1}$.  indicate the
importance of intermediate asymptotic regimes, as discussed in the
main text Sec.~''Crossovers and Beyond Asymptopia'' (main text). 

While we have focussed on the marginal case near $\mu=d$ in this
article, it is clear that another case of marginal stability controls
the cross-overs near $\mu=d+1$. Simulation results reported in
supplementary Fig.~\ref{fig:second-marginal-point} indicate that
$\ell(t)/t\sim \log(t)$ for $\mu=2$ in one dimension. This is
consistent with our funnel argument: With nearly constant speed, the
gap between the funnels remains roughly constant for a time of order
$t$. To ensure that the source emits about one jump to the target
funnel, we must have that, per unit time, the probability of jumping
over the gap is of order $1/t$. Thus, the gap size $\Delta \E$ should
be such that $\Delta \E^{-\mu}\sim 1/t$. For $\mu=2$, we thus have
$\Delta \E\sim t$, i.e., the key jumps span distances of order
$t$. This is ensured when $(\ell(t)-2\ell(t/2))/t\sim$ const., i.e.,
if $\ell(t)/t\sim\log t$. Note that a rigorous upper bound of this
form follows from the arguments presented in
Sec.~\ref{Sec:intermediate-range-case} for the regime $\mu>d$.  The
jumps of order $O(t)$ that drive the logarithmic increase in spreading
velocity might be the ``leaps forward''~\cite{mollison1972rate}
recognized by Mollison in one of the earliest studies on spreading
with long-range jumps.

\subsection{Occupancy profiles and relevance of secondary seeds}
\label{sec:dens-prof-relev}
In the main text, we introduced the notion of a nearly occupied core
of the population (of size $\ell(t)$) as the source of most of the
relevant seeds in the target funnel. However, it is clear that outside
of this core there is a region of partial occupancy. This region is
potentially broad, in particular for $\mu\to 0$, and may therefore
lead to a significant fraction of relevant seeds. An improved theory
should account for those secondary seeds, and should also be able to
determine the profile of mean occupancy or, equivalently, the
probability that a site is occupied. While we give rigorous bounds on
how the total population grows in Sec.~\ref{sec:rigorous}, we first use an
improved version of our funnel argument to describe the occupancy
profiles. 

We focus on the probability, $\q(r,t)$, that a time $t$ after a mutant
establishes, it will have taken over the population a distance $r$
away.  We expect that $\q(r,t)$ will be close to unity out to some
core radius $\ell(t)$, and then decrease for larger $r$, with the
average total mutant population proportional to $\ell(t)^d$.  With
only short-range dispersal, $\ell(t)\approx vt$ and the core is
clearly delineated but when long jumps are important the crossover from
mostly occupied core to sparsely occupied halo will not be sharp. The
more important quantity is the average of the total area (in
two-dimensions or linear extent or volume in one or three) occupied by
the mutant population, we denote this $\M(t) =\int d^dr \q(r,t)$. 

To find out when long jumps {\it could} be important,  we first ask
whether there are likely to be {\it any} jumps longer than $\ell(t)$
that occur up to time $t$.  The average number of such long jumps is
of order $t\ell(t)^d  \int_{\ell(t)}^\infty  r^{d-1}dr G(r) $.  If
$G(r)$ decreases more rapidly than $1/r^{2d+1}$, this is much less
than $t/\ell(t)$ for large $t$.  As $\ell(t)$ increases at least
linearly in time, the probability that there have been any jumps
longer than $\ell(t)$ is very small. The guess that $\ell$ indeed
grows as $vt$, and consideration of  jumps that could advance the
front fast enough to contribute substantially to $v$, leads,
similarly, to the conclusion that there is a maximum $t$-independent
jump length beyond which the effects of jumps are negligible, indeed,
their effect decreases more rapidly than $G(r)$. This reinforces the conclusion
that there is only linear growth with $\mu>d+1$:  a very strong
breakdown of the deterministic approximation which yielded exponential growth for any power law.

When $G$ is longer range, in particular if $G(r) \sim 1/r^{d+\mu}$
with $\mu<d+1$, many jumps longer than $\ell(t)$ will have occurred by
time $t$.  We now investigate the effects of such long jumps on the density profile.  To do
so, we investigate the behavior of $\q(R,T)$ for large $R$ and $T$ in
terms of the $\{\q(r,t)\}$ at shorter times and --- primarily ---
corresponding distances $r \sim \ell(t) $ which can be much less than
$R$.  Mutants can get to a chosen point, $\bR$, by one making a long
jump at time $t$ from a starting point $\bx$ to an end point, $\by$,
and subsequent spread from there to $\bR$ during the remaining time
interval of duration $T-t$.  The rate (per volume elements) of this
occurring is $\q(\x,t)G(|\bx-\by|) \q(|\bR-\by|,T-t)$. In the
approximation that these are independent, the probability that this
does not occur at any $t<T$ from any $\bx$ to any $\by$ is simply
Poisson so that 
\be \q(R,T)\approx 1- e^{-\Q(R,T)} \label{eq:occupancy-probabili} \ee with \be
\Q(R,T) \approx \int_0^T dt \int_{z>\ell(T/2)} d^d\z \int d^d\x\,
\q(\x,t)\,G(z) \,\q(|\bR-\bx-\bz|,T-t) \ .  \label{eq:Q-definition}
\ee Here, we substituted the final point $\by=\bx+\bz$ by the sum of
the jump start site and a jump vector $\bz$, over which we integrate.
Note that a lower cut-off in the $z$ integral is necessary to exclude
the many very short jumps that lead to strongly correlated
establishments. The cut-off is also necessary to not count mutants
that results from growth in the target area, rather then seeding from
the source funnel. Our main assumption here is that if a single seed
is sufficiently far from other seeds or occupied regions then the 
growth from the seed  is independent of the rest of the system as long as collisions are
unlikely. 


When the jump integral is strongly peaked at $z=R$, as is the
case in or close to the stretched exponential regime, the final
results will be independent of this cutoff to leading order. Then, we
can approximate $Q(R,t)$ as
\begin{equation}
  \label{eq:Poisson-variable}
  Q(R,t)\approx G(R)\int_0^{t}dt' M(t') M(t-t') \;,
\end{equation}
where $M(t)$ is the expected total size of a population at a time $t$,
\begin{equation}
  \label{eq:mass}
  M(t)=\int d^dx  q(x,t) \;.
\end{equation}

For a power-law kernel $G(R)=G(1)R^{-(1+\mu)}$, we make the ansatz
that Eqs.~\eqref{eq:Poisson-variable},\eqref{eq:mass} and
\eqref{eq:occupancy-probabili} can be approximately solved by a scaling form
\begin{eqnarray}
  \label{eq:scaling-form-densities}
  q(R,t)&=&\Xi\left(\frac{R}{\lambda(t)}\right) \\
 {\rm with} \ \  \Xi(\xi)&=&1-\exp\left (-\xi^{-(d+\mu)}\right)\;,
\end{eqnarray}
which leads to the condition
\begin{equation}
  \label{eq:Q-scaled}
  Q(\xi\lambda(t),t)=\xi^{-(d+\mu)} \kappa^2 G(1)
  \lambda^{-(d+\mu)}\int_0^t dt' \lambda^d(t') \lambda^d(t-t') \;,
\end{equation}
where $M(t)=\kappa\lambda(t)^d$ and $\kappa_\mu$ is given by
\begin{equation}
  \label{eq:kappa}
  \kappa=\int d\xi^d \Xi(\xi)\;.
\end{equation}
Thus, the above scaling form is a valid solution if the characteristic
scale $\lambda(t)$ satisfies
\begin{equation}
  \label{eq:lambda}
  \kappa^2 G(1)
  \lambda^{-(1+\mu)}\int_0^t dt' \lambda(t') \lambda(t-t')=1 \;.
\end{equation}
The resulting condition is similar to our condition in the main text
but differs by the numerical factor $G(1)\kappa_\mu^2$. In one
dimension,
\begin{equation}
  \label{eq:kappa-1d}
  \kappa_\mu=2\Gamma\left(\frac{\mu}{1+\mu}\right) \;.
\end{equation}
The divergence $\kappa_\mu^2 \sim \mu^{-2}$ as $\mu\to0$ indicates the
importance of seeds from the tail regions for small $\mu$.


Notice that at long distances, $R\gg\ell(T)$, the length of the jumps
that dominate are close to $R$ so that our approximation for $\Q(R,T)$
should be correct even if it is a poor approximation for $R\approx
\ell(T)$.  Thus the decrease in $\q$ at large distances is simply
proportional to $G(R)$, more specifically, 
\be
\q(R,T) \sim \frac{G(R)}{G(\ell(T))}
\ee
 (as predicted by the scaling form) so that it is of
order unity at $R\sim \ell(T)$.  Note that this implies that, since
$G(r)$ is integrable, $\int d^dr \q(r,t)$ is indeed dominated by
$r\sim \ell(t)$ as we have assumed. This long distance form for the density profile is also found in the analyses of upper and lower bounds in the next sections: thus it can be readily proved along the same lines. 


\section{Rigorous bounds}
\label{sec:rigorous}

As most of our results are based on approximate analyses and heuristic arguments, it is useful to supplement these by some rigorous results.   We focus on the one-dimensional case: extensions to higher dimensions can be done similarly, although with a few complications that will require some care.  We sketch here the arguments that can lead to proofs without all the details filled in.  

As via the heuristic arguments, we would like to obtain the behavior at longer times in terms of the behavior at shorter times, in particular times around half as long.  

We would like to prove that there exist time dependent length scales, $\LL(t)$ and $\LG(t)$  and functions, $\FL(r,t)$ and $\FG(r,t)$ such that the probability, $\q(r,t)$, that a site at $r$ from the origin is occupied at time $t$, is bounded above and below by  
\be
\FL(r,t) < \q(r,t) < \FG(r,t)  \ \  {\rm with} \ \  M_< \equiv \int dr \FL(r,t) \propto \LL(t) \ \  \& \ \  \M_> \equiv \int dr \FG(r,t) \propto \LG(t) 
\ee
for all times.    Then $\LL$ and $\LG$ are lower and upper bounds for  $\ell(t)$ --- with some appropriately defined definitions of $\ell(t)$ which differ somewhat, although not significantly,   for the upper and lower bounds. 

While we  would like the upper and lower bounds on $\ell(t)$  to be as close as possible to each other,  in practice, we have obtained bounds that are good on a logarithmic scale: i.e. for $\log \ell(t)$, rather than on a linear scale.   

Similarly, we would like to have the bounds be close to the actual expected form of $\q$, with $\FL$ very close to unity for $r \ll \LL$ and proportional to $\lb[\frac{\LL}{r}\rb]^{\mu+1}$ for $r \gg \LL$ and similar for the upper bounds. 
It is often more convenient to consider the typical time-to-occupation as a function of the distance, $\tau(r)$ and derive upper and lower bounds for this, $\TG(r)$ and $\TL(r)$, respectively, 
such that 
\be
\LL(t=\TG(r))=r  \ \ \ \& \ \ \ \LG(t=\TL(r))=r
\ee
with 
\be
\TG(r) > \tau(r) > \TL(r) \ .
\ee
Because of the faster-than-power-law growth, bounds on $\tau(r)$ are generally much closer than those on $\ell(t)$. 

As we would like to justify the use of the heuristic iterative scaling arguments more generally, it is especially useful to obtain iterative bounds directly of the form used in those heuristic arguments: $\ell(T)$ in terms of $\{\ell(t)\}$ for $t$ in a range near $T/2$.  
As the heuristic arguments do, in any case, only give $\ell(T)$ up to a multiplicative coefficient of order unity, we will generally ignore such order-unity coefficients in length scales except for coefficients that  diverge or vanish exponentially rapidly as $\mu \to d$, in particular in the intermediate range regime the coefficient, $A_\mu$ in $\ell(t)\sim A_\mu t^{1/(\mu-d)}$  which vanishes  as $\log (A_\mu) \approx - \log 4 /(\mu-d)^2$  as $\mu \searrow d$.

\subsection{Upper bounds}

\subsubsection{Simple power-law bound}

The simplest bound to obtain is an upper bound for $\ell(t)$ in the short and  intermediate range regimes: i.e.,  in one dimension, $\mu >1$.   Define $\E(t)$ to be the right-most edge of the occupied region at time $t$, i.e. $\c(x,t) =0$ for $x>\E(t)$.   The probability of a jump that fills a position $y>E(t)$ in $(t,t+dt)$ is less than $\int_{-\infty}^{\E(t)} dx G(y-x) \sim 1/(y-\E(t))^\mu$.  For $\mu>1$, the lower extent of the integral can be taken to $-\infty$ as the jumps arise, predominantly, from points that are not too far from the edge. 
[In contrast,  for $\mu<1$ jumps from the whole occupied region are important and  this bound would yield a total jump probability to long distances that diverged when integrated over $y$, and one would have to instead use a  lower extent of the integral of $-\E(t)$ for the left edge.]

The advancement of the edge is bounded by a translationally and temporally invariant process of jumps of the position of the edge by distances, $\Delta\E$, whose distribution has a power-law tail.   For $\mu>2$, the mean $\langle \Delta\E \rangle <\infty$, implying that the edge, and hence $\ell(t)$,  cannot advance faster than linearly in time.  
But for the intermediate regime, $\langle \Delta\E \rangle =\infty$ so that $\E(t)$ could advance as fast as a one-sided Levy flight with $\E(t) $  dominated by the largest advance.  As this process would  yield $\E \sim t^{1/(\mu-1)}$, this  implies that $\ell(t)$ is bounded above by the same form as the heuristic result. 

Although the simple bound captures some relevant features, in particular the dominance of jumps of length of order $r$ to fill up a point at distance $r$, it is otherwise rather unsatisfactory. First, the coefficient does not vanish rapidly as $\mu \searrow 1$. And second,  it suggests that the probability that an anomalously distant point, $r\gg\ell(t)$, is occupied, is, in this crude approximation of full-occupancy out to the edge, simply the probability that $\E(t)>r$ which  falls-off only as $1/r^{\mu-1}$ --- much more slowly than the actual $\q(r,t) \sim r^{-1-\mu}$. 

Nevertheless, for proving better upper-bounds, the Levy-flight approximation for the dynamics of the edge is quite useful. 

\subsubsection{Upper bounds from source-jump-target picture} 

As discussed earlier, one would like to make the heuristic argument of a single long jump from a source region to a target funnel region include also --- or provide solid reasons to ignore ---  the effects of jumps from the  partially filled region outside the core of the source. Very loosely, one would like to write the probability  that a point, $R$,  is not occupied at time $T$, as 
\be
1-\q(R,T) \approx \exp\lb[-\int_0^T dt \int_{-\infty}^\infty dx  \int_{-\infty}^\infty dy \q(x,t) G(|y-x|) \q(R-y,T-t)) \rb] 
\ee
with $x$ in the source region and $y$ in the funnel of $R$.    But for any positive $\mu$, the spatial integral is dominated by $y-x$ small, so that this does not properly represent the process: there is a drastic over-counting of short jumps. 

One can do much better by trying to separate the long jumps from the short ones, and the source region from the funnel  (in the crude approximation these overlap).   To do this, we choose, for the $R$ and $T$ of interest, a spatio-temporal source region, $S$, around the origin which has a boundary at distance $\BS(t)$  that loosely reflects the growing source: $d\BS/dt \ge 0$.   We then separate the process of the set of jumps that lead to $R$ into three parts.  First, jumps solely inside $\S$ which lead to a spatio-temporal configuration of occupied sites, $\{\cs(x,t)\}$; second, {\it bridging} jumps from these out of $\S$, say  at time $t$ from $x$ in $S$ to a point $y$ in the rest of space-time, $\bar{\S}$,  and third all the subsequent dynamics from such seeds in $\bar{\S}$ including inside and outside $\S$ and between these.   This over-counts the possible spatio-temporal routes to $R,T$ --- especially as returns to inside $\S$ from outside are included --- and  thus provides an upper bound for $\q(R,T)$.  The probability, $p_a$,  that a single seed, $a$,  to $y_a$ at $t_a$ leads to $R$ being filled by $T$ is  $\q(R-y_a, T-t_a)$.  But the probability that a second seed, $b$, leads to $R$ filled by $T$ is not independent as the fate of these seeds involves overlapping sets of jumps: indeed, they are positively correlated so that 
\be
\pr[\c(R,T) = 0 | seeds \ a \ \& \ b] \ge \pr[\c(R,T) = 0 | seed \ a ] \times \pr[\c(R,T) = 0 | seed \  b]  \ .
\ee
Since for a given $\{\cs(r,t)\}$, the probability density of a seed at $y,t$  is $dydt \int_{|x|<\BS(t) }\cs(x,t)G(y-x)$,  and using the generalization of the above bound to many seeds, we have
\be
\q(R,T) \le 1- \exp\lb[-\int_0^T dt \int_{|x|<\BS(t)} dx\,  \qs(x,t) \int_{|y|>\BS(t)} dy\, G(y-x) \q(R-y,T-t))\rb]  \label{q-up-iter}
\ee
where $\qs(x,t)\equiv\langle \cs(x,t) \rangle$ and we have used $\langle \exp(X) \rangle \ge  \exp(\langle X \rangle)$ for any random variable.     

To derive a useful upper bound on $\q(R,T)$ we need to choose appropriately the boundary, $\BS(t)$, of the source region and put a sufficiently stringent upper bound on $\qs(x,t)$.    

\subsubsection{Long-range case}

For the long-range case, $\mu<1$,  the integrals over $x<\BS$ and $y>\BS$ of $G(y-x)$ 
are dominated by long distances. Thus the short jumps from inside to outside $\S$ do not contribute significantly.  We can then simply replace $\qs$ by the larger $\q$ to obtain a slightly weaker bound which is of exactly the form of the naive estimate except for the strict delineation of the source region which prevents the most problematic over-counting of the effects of short jumps.    
 A  particularly simple choice is  $\BS =\frac{1}{2}R$ independent of $t$. 

We now proceed by induction.  Take the bound on the scaling function to have the form  $\FG(r,t)=1$ for $r<\LG(t)$ while $\FG(t)= \lb[\frac{\LG}{r}\rb]^{\mu+1}$ for $r \gg \LG(t)$ and assume that for some appropriate $\LG(t)$, this is indeed an upper bound  for all $t<T$: i.e., $\q(r,t) \le \FG[r/\LG(t)]$.    We can now use \eqref{q-up-iter} with $\qs$ and $\q$ both replaced by $\FG$.  

When $\LG(t)$ and $\LG(T-t)$ are both much less than $R$,  the integrals over $x$ and $y$ will be dominated by the regions near the origin and $R$ respectively, yielding  the spatial convolution $ \FG \circ G \circ \FG \sim \frac{\LG(t)\LG(T-t)}{R^{\mu+1}}$. There are small positive corrections to this from two sources: first, from the regions near  $0$ and $R$, which, by expanding $y-x$ in $x$ and $R-y$, are seen to be of order $ [\LG(t)^2+\LG(T-t)^2]\LG(t)\LG(T-t)/R^{\mu+3}$;  and, second,  from $y-x \ll R$, the region near the source boundary, which are of order $\lb[\frac{\LG(t)}{R}\rb]^{\mu+1}\lb[\frac{\LG(T-t)}{R}\rb]^{\mu+1}R^{1-\mu}$ with the last part from the integrals over $x$ and $y$.  As the dominant part  is exactly of the form in the heuristic treatment, integrating it over time is strongly peaked at $t\approx T/2$ (note that for either $t$ or $T-t$ much smaller than $T$,  one of the $\FG$ factors will be close to unity near the boundary, but these ranges of time only contribute weakly). If we use $1-e^{-Q} \le \min(Q,1)$ then $\LG(T)$ can be chosen as the value of $R$ for which $Q=1$, and for $R\gg \LG(T)$ the desired $\FG \sim \lb[\frac{\LG}{R}\rb]^{\mu+1}$ is obtained. Including the small correction factors in the convolutions necessitates slight modifications of the recursion relations for $\LG$ but these are negligible at long times.   

\subsubsection{Intermediate-range case}
\label{Sec:intermediate-range-case}
Obtaining an upper bound in the intermediate range case is somewhat trickier.  If we again replaced $\qs$ by $\q$, then the integrals over $x$ and $y$ would have a part dominated by both points being near the boundary: with $\BS \sim R$, this contribution to the convolution would be of order $\lb[\frac{\LG(t)}{R}\rb]^{\mu+1}\lb[\frac{\LG(T-t)}{R}\rb]^{\mu+1}$. With $t \sim T/2$ and $R\sim \ell(T)$, all the lengths should be of order $t^\beta$, so that this boundary piece is larger by a factor of $[\ell(T)^\beta]^{\mu-1} \sim T$ than what-should-be the dominant part from $x$ and $y$ near $0$ and $R$, respectively.   Thus we need a better upper bound on the restricted-source $\qs(r,t)$ which vanishes as $r \nearrow \BS(t)$. 

To bound $\qs$, we can make use of the simple bound for the edge of the occupied region derived above, combined with the restrictive effects of the boundary, $\BS(t)$. Instead of choosing $\BS$ to be constant, we choose it to have constant slope, $U\equiv d\BS/dt$,  of order $\ell(T)/T$.      As jumps that contribute to $\qs$ are not allowed to cross the boundary,   the distribution of jumps of the edge $\E(t)$ is cutoff at $\zeta(t)\equiv\BS(t)-\E(t)$.  Since $U>v_0$,  the speed of spread in the absence of jumps beyond nearest neighboring sites, typically the gap, $\zeta(t)$, will increase with time, decreasing only by jumps.  The sum of all the jumps of $\E$  in a time interval $\Delta t$ is dominated by the largest, which is of order $(\Delta t)^{1/(\mu-1)}$. This would result in the edge moving faster than $U$ except for the cutoff. The typical gap, $\tilde\zeta$, is then obtained by balancing its steady decrease against the dominant jump: $U\Delta t \sim (\Delta t)^{1/(\mu-1)}$ yielding $\Delta t \sim U^{(\mu-1/)(2-\mu)}$  and hence
\be
\tilde{\zeta} \sim U^{1/(2-\mu)} \sim \lb[\frac{\ell(T)}{T}\rb]^{1/(2-\mu)} \sim \ell(T) A_\mu^{(\mu-1)/(2-\mu)}
\ee
using $\ell(t) \sim  A_\mu t^{1/(\mu-1)}$.  In the limit of $\mu \searrow 1$,  $\tilde{\zeta}/\ell \sim 4^{-1/(\mu-1)}$, vanishing rapidly --- a reflection of the strong failure of the simple edge-bound in this limit but sufficient for our present purposes.  In a time $\Delta t \ll T$, the distribution of $\zeta$ in this approximation will reach a steady state.   The probability that $\zeta \ll \tilde\zeta$ is controlled by the balance between jumps of $\E$ to near the boundary, and the steady increase in $\zeta$ from the boundary motion: its probability density is hence of order $\zeta/\tilde\zeta$ which, since in this approximation all sites are occupied up to $\E$, implies that $\qs$ vanishes at least quadratically for small gap $\zeta$. Combining this with the trivial bound of $\qs< \q$ and choosing a convenient normalization of $\tilde\zeta$, we thus have
\be
\qs(r,t)  \le  \min\lb[\FG(r,t), \frac{(\BS(t)-r)^2}{\tilde{\zeta}^2}\rb] \ .  
\ee
It remains to chose $\BS(t)$ so that the bound on $\qs$ remains sufficiently good for $t$ small enough that  the steady state distribution of $\zeta(t)$ has not yet been reached.  To keep $\E(t)$ typically of order $\tilde\zeta$ from $\BS(t)$, we can simply chose $\BS(0)=\tilde\zeta$ and $U= (R-2\tilde\zeta)/T$. 

With our improved bound on $\qs$, for  the convolution $\qs \circ G \circ \q$, the small $\BS-x$  parts are no longer dominated by $\BS-x$ of order unity, but by $\BS-x$ near the crossover point between the two bounds on $\qs$. This yields a contribution to the convolution of order $\LG(t)^{\alpha_S}\LG(T-t)^{\alpha_F} / R^{\alpha_S+\alpha_F +\mu -1}$ with $\alpha_F=(\mu+1)(2-\mu)$ and $\alpha_S=\alpha_F (3-\mu)/2$ and  a multiplicative coefficient that does not depend exponentially on $1/(\mu-1)$ because the integral over $x$ scales as $1/\tilde{\zeta}^{\mu-1}$.
As $\mu \searrow 1$,  $\alpha_F\to \alpha_S\to1$, and the boundary contribution is less than the dominant part uniformly in $t$.
Note that for $\mu>\mu_B \cong 1.5$, the bound on the near-boundary contribution can be somewhat larger for $t<T/2$ than the dominant parts, but it scales in the same way with $T$ and thus only weakens the upper bound on the coefficient, $A_\mu$, which is in any case of order unity in this regime.   

Once the over-counting of short jumps has been sufficiently reduced, as we have now done, the rest of the analysis, in particular the large $R/\LG$ form of $\FG$, follows as in the long-range case.

The {\it marginal case} $\mu=1$ can be analyzed similarly to the intermediate-range case, resulting in an additional logarithmic dependence on $R$ of the near-boundary contribution which is, nevertheless, still much smaller than the dominant part. 

The upper bounds that we have obtained are, except for modifications at small scales and for $\mu$ not much smaller than 2, essentially the same as given by  the heuristic arguments, thus only differing at long scales by order-unity coefficients which, in any case, we did not expect to get correctly.  All the crossover behavior near $\mu=1$ is in the upper bounds, although that near $\mu=2$ is not.

\subsection{Lower bounds}

To obtain lower bounds on the growth of the characteristic length scale $\ell(t)$ and the occupation probability, $\q(r,t)$, a different strategy needs to be employed.
One of the difficulties is the dependence on the behavior at each time scale on all the earlier time scales: for the filled region to grow typically between time $T$ and $2T$,  the stochastic processes  that lead to the configuration $\c(x,T)$ must not have been atypically slow or ineffective.  As this applies iteratively scale by scale,  we must allow for some uncertainty in whether or not the smaller scale regions are typical, leading to some uncertainty at all scales which, nevertheless, we need to bound. Because of the stochastic heterogeneity of $\c(x,t)$  it is better to focus on a coarse-grained version of the occupation profile rather than on $\c(x,t)$ itself, as integrations over $\c$ at time $t$  are what act as the sources of future occupation at larger distances. 

\subsubsection{Mostly-filled-in: marginal and long-range regimes}

We consider the probability that a region is almost full,  in particular, with a seed at the origin,  we consider the region to one side of the origin and define
\be
\PF(r,t;\Phi) \equiv \pr\lb[\frac{1}{r}\int_0^r dx\, \c(x,t) > \Phi\rb]
\ee
with $\Phi$ close to or equal to unity being of particular interest.  
In order to keep events sufficiently independent, we consider, as for the upper bounds, the probability of events that do not involve any jumps out of some region.  In particular, defining $\PS(r,t,\Phi)$ similarly $\PF$, but with the restriction that jumps  do not go out of the interval  $(0,r)$.  
For the long-range and marginal cases, we will focus on partial filling, but  for the intermediate range case the scale invariance mandates different treatment so we instead analyze  full filling -- i.e. $\Phi=1$. 

The basic strategy is to start with a particular deterministic approximation to $\ell(t)$, $\LT(t)$ with corresponding times $\TT(r)$, and then show that at time not  too large a multiple of  $\TT(r)$, the region out to $r$ will be nearly filled  with high probability: i.e. that $\PS(r,\TG(r);\Phi)$ is close to unity for $\TG(r)/\TT(r)$ sufficiently large.  We will be interested in large scales as, in any case, fluctuations at the  small scales can only change coefficients by order unity. We can thus be sloppy with some of the bounding inequalities: these could  be improved to include the ignored corrections to the large scale effects to make fully rigorous bounds. 

As the range of time over which the typical $\ell(t)$ expands
significantly plays an important role, it is useful to define \be
\DT(t) \equiv \lb [ \frac{d\log \LT(t)}{d\log t}\rb]^{-1} \ee which is
small except for $\mu$ substantially larger than one.  The dominant
jumps from source to funnel involve an integral over time of
$\LT(t)\LT(T-t)$, which is primarily from a range of order
$T\sqrt{\DT}$ around $T$ as discussed above.  The
deterministic-iterative approximation that we use as a base for the
lower bounds is the solution to the iterative relation (ambiguous up
to an ${\cal O}(1)$ multiplicative factor which we ignore throughout):
\be [\LT(2T)]^{\mu+1}= T[\LT(T)]^2\sqrt{ \DT(T)} \ee corresponding to
roughly one seed into a funnel of width $\LT(T)$ from a jump of
distance $\LT(2T)$ from the source up to time $T$.  For convenience,
we use only half the source --- $x$ from $0$ to $\LT(T)$. The results,
$\LT(t)$, of this iterative approximation are, up to numerical factors
that arise from these modifications and from other from non-asymptotic
effects at small scales, equivalent to the upper bounds, $\LG(t)$ from
the above.  In particular, the ratio between the corresponding times,
$\TT(r)$ and $\TL(r)$ we expect to approach constants that are not
singular near the marginal case $\mu=d=1$.

For the lower bounds  it is convenient to work with a specific set of length scales, $\LT_n=\LT(\TT_n)$, corresponding to a series of time scales, $\TT_n=2^n$ (dropping a prefactor). 
To mostly fill out to $\LT_{n+1}$ without jumps going out of $(0,\LT_{n+1})$ from the source of size $\LT_n$, most of the $\K_n = \LT_{n+1}/\LT_n $ bins of size $\LT_n$ must be mostly filled.  To get a  lower bound on how long this takes and how likely it is, we make several simplifications each of which lead to underestimates of the probability that the desired filling has occurred.  First, consider only jumps into each bin that  come directly from the source (rather than from other bins as can occur later). Second, ignore all but the first seed jump  from the source into the bin  (the effects of later jumps  are not independent of those of the first). And third,  include only jumps that lead from the seed in a bin that do not go outside that bin during the time during which  the probability of it being mostly filled is considered. 
The last two conditions mean that the probability that the bin is filled to a fraction $\Phi$ by a given time, $t$, after   
the seeding jump, is at least as large as $\PS(\LT_n,t;\Phi)$  since a seed at the edge of the bin, which corresponds to the definition at the source, is less likely to mostly fill the bin than a seed away from the edge. 

At large scales for $\mu\le 1$, the number of bins, $\K_n$, grows with scale: $\K_n \sim \sqrt{\TT_n}$ for the marginal case and larger for the long-range case.  Thus if the probability that the furthest bin from the source is mostly filled is $f_n$, with  the filling of the others being more probable  as they are closer,  it is likely that the number that are similarly  mostly  filled is close to $\K_n f_n$, with significant deviations from this being  very unlikely at large scales.  In order to iterate while not losing too much in filling fraction, we chose a series of partial filling fractions, $\{\phi_n\}$, such that $\Phi_N\equiv\prod_{n=1,N-1} \phi_n $ converges to  the desired overall filling fraction, $\Phi$, at large $N$, and chose conditions such that $f_n$ is sufficiently large that the fraction of the $\K_n$ bins filled to $\Phi_n$ is greater than $\phi_n$ with high probability: this then implies that  the region from the origin to $\LT_{n+1}$ will be filled to greater than $\Phi_{n+1}$ with high probability.  A convenient choice is $\phi_n = 1- \Delta/n^{1+\alpha}$ with any positive $\alpha$ and $\Delta \sum_n n^{-1-\alpha} < 1-\Phi$.  For convenience in dropping $\log\Delta$ factors that otherwise appear in many places, we restrict consideration to $\Delta$ not very small, and do not keep careful track of $\alpha$ factors that also appear as one can take $\alpha \to 0$ at the expense of corrections that are down by one extra logarithm.   

The filling probability of a bin is at least as large as that obtained from the requirement of the occurrence of  both of two independent events: a jump into the bin from the source that occurs before some chosen initial time, $T_I$, and the bin being filled from that single seed by a time, $T_B+T_I$. The probability of a jump into a bin is at least $1-e^{-W_n}$ in terms of a conveniently chosen lower bound, $W_n$, on the expected number of jumps from the source into the furthest away bin, and the probability of the bin being filled from the single seed  is at least $\PS(\LT_n,T_B;\Phi_n)$.  We will find iterative bounds on $\PS$ that it is convenient to write in the form 
\be
\PS(r,t;\Phi) \ge 1 - e^{-\Lambda(r,t;\Phi)}
\ee
so that 
\be
1-f_n \le e^{-W_n} + e^{\Lambda_B}  \ \ \ {\rm with} \ \ \ \Lambda_B\equiv\Lambda(\LT_n,T_B;\Phi_n)\ .
\ee
For convenience we chose conditions so that $\Lambda_B \ge W_n$ and $1-f_n \le \frac{1}{2}(1-\phi_n)$ which, for $K_n$ large, makes the probability that a fraction $\phi_n$ of the bins are not filled exponentially small. We henceforth ignore this factor in the probability as it does not matter except on small scales: adjustments to take it into account are straightforward.  We thus require that 
\be
\Lambda_B \ge W_n \ge \log[ (1-\phi_n)/4] =  (1+\alpha) \log n  + {\cal O}(1) \ .
\ee

To obtain a bound on $\PS(\LT_{n+1},T;\Phi_{n+1})$, we must show that a source that can give rise to an average number at least $W_n$ of jumps into the furthest bin by time $T_I$, occurs with probability that is somewhat larger than the desired bound at the next scale.  The expected effective number of jumps out of the source of size $\LT_n$  into a bin of the same size a distance up to $\LT_{n+1}$ away before time $\TT_n$ was assumed in the deterministic iterative approximation to be of order $\sqrt{\DT_n}\TT_n$.  In order to ensure that the average number from the actual source is sufficiently large,  we can require that it be almost filled by some time $T_S$ and only include jumps that occur between $T_S$ and $T_I$ as the rate of these is bounded below by the filling at $T_S$. The required range is  \be
T_I-T_S=W_n\TT_n\sqrt{\DT_n}/\Phi_n \ .
\ee

We now proceed by induction and show that if 
\be
\Lambda(\LT_n,t;\Phi_n) \ge \gamma_n(t-U_n\TT_n)
\ee 
for $t$ in a range such that $\Lambda$ is relatively large --- the precise range is not crucial but minor modifications are needed to extend out to arbitrary large $t$ --- then a similar bound holds at the next scale with coefficients $\gamma_{n+1}$ and $U_{n+1}$ with both these varying slowly with $n$ at large scales.  
Note that at the smallest scale the probability that a site is filled by a jump directly from the origin by time $t$ converges exponentially to unity for long $t$, thus at the smallest scales there is a trivial bound of this form. As the scale is increased, $\gamma_n$ will initially change, but once the scale becomes large enough that the width of the distribution of the fraction of the bins mostly filled is small, then $\gamma_n$ saturates and becomes weakly dependent on $n$. In the analysis below, it can be replaced by a constant. 

Consider a total time $T$ to mostly fill out to $\LT_{n+1}$.  The time for the bins to fill with sufficiently high probability once they have been seeded is $T_B\le U_n\TT_n +W_n/\gamma_n$. With $T_I -T_S$ as above, we have a time for the source to fill
\be
T_S \ge T - \TT_n[U_n +W_n\sqrt{\DT_n}/\Phi_n)] -W_n/\gamma_n \ .
\ee
Plugging in the probability that the source is filled in this time gives a bound on $\Lambda(T, \LT_{n+1},\Phi_{n+1})$ of the same form but with, dividing out $\TT_{n+1} = 2\TT_n$, 
\be
U_{n+1} \le U_n + \frac{W_n\sqrt{\DT_n}}{2\Phi_n} +\frac{W_n}{2\gamma \TT_n} \ .
\label{U-itern}
\ee
As $\TT_n$ increases rapidly and $W_n$ only slowly, the last term only contributes at small scales.  

For the {\it long-range regime}, $\DT_n \sim e^{-\eta \log 2 n}$ so the second term in \eqref{U-itern} is also small except at small scales and we conclude that $U$ is bounded above by a $\mu$ dependent constant.  Thus the lower bound for $\ell(t)$, and upper bound for $\tau(r)$ have exactly the same form as the opposite bounds, except with the scale of $t$ --- i.e. $B_\mu ^{-1/\eta}$ --- different.  

For the {\it marginal case}, $\DT_n\approx \frac{2}{n}$ so that $U_n$ changes slowly at large scales.  Integrating up, one sees that 
\be
U_n < C\sqrt{n}  \log n \sim \sqrt{\log \TT_n} \log\log \TT_n  
\ee
with a coefficient independent of $n$ (but depending on $\Phi$ and $\alpha$). 
One can now solve for the time scale above which mostly filled is likely, $\TG(r)=U(\TT(r))\TT(r)$, to find a lower bound, $\LL(t)$, on $\ell(t)$:
\be
\log(\ell(t))\ge \log(\LL(t))= \frac{\log(t)}{4\log 2}[\log t - 2 \log\log t - {\cal O}(\log\log\log t)]  \label{low-bnd-marginal}
\ee
which is very close to the upper bound derived above,
\be
\log(\ell(t))\le \log(\LG(t))= \frac{\log(t)}{4\log 2}[\log t -  \log\log t - {\cal O}(1)]   
\ee
differing only in the coefficient of the correction term.

One of the advantages of this iterative approach is that the {\it crossover regime} can be handled similarly by integrating up \eqref{U-itern}.  The lower bound will be similar to the upper bound throughout this crossover regime and into the asymptotic regimes for the marginal and long-range cases.  

\subsubsection{Fluctuations and intermediate-range regime}

The reason that the fluctuation effects are relatively small for the marginal and long-range regimes is that at each successive time scale, more and more roughly independent long jumps are involved in filling up to the next length scale: i.e. $K_n$ continues to grow.  For the long-range regime, it grows so rapidly that almost all the fluctuations come from early times: this is like what occurs for the fully mixed model.   For the marginal case, the fluctuations are dominated by the smallest scales but the cumulative effects of them over the longer scales does make a difference as found in obtaining the lower-bounds.  

For the intermediate-range case, the ratio of length scales for each factor of two in time  scales saturates (when out of the crossover regime) at $K\approx 2^\beta$. This means that whatever fraction, $\phi_n$, of the bins are to be filled at each scale,  the probability that this occurs either decreases with scale if the product of the $\phi_n$ does not go to zero, or saturates to a constant if the $\phi_n$'s do also, in which case the overall filling fraction $\Phi_n$ tends to zero as a power of time.  
At each scale there are a comparable number of long jumps that are needed, thus one should expect that fluctuation effects will be  scale invariant and not decrease with scale. 

To get useful lower bounds on $\ell(t)$ via an upper bound on $\tau(r)$, $\TG(r)$, the easiest way is to require that the source be completely full, and that jumps from this completely fill all the bins at the next scale: this avoids the problems with the $\phi_n$.  As the probability that all the bins are filled is  only (readily) bounded by $ (1-e^{-W} -e^{-\Lambda_B})^K\approx 1-2Ke^{-W}$ if we  again chose $\Lambda_B \ge W$, $W$ must be larger by $\log K \approx \beta \log 2$ than for the partially filled analysis above.   Carrying through similar analysis, gives  for large $\beta$ a coefficient $\gamma_n \approx 2\sqrt{\beta}/(n\TT_n)$ which means that the filling probability decays for large times as roughly the inverse of the typical time --- natural as the needed long jumps that occur at rate $\sim 1/\TT_n$  have a distribution of when they occur on the same time scale.  Note the contrast to the rapid decay of the not-mostly-filled probability on a time scale of order unity from the partially filled bound derived above. 
The time beyond which the full filling is likely is only bounded by, in this analysis,  
$\TT_n U_n \sim \TT_n \sqrt{\beta} n^2 $.  This gives a lower bound on $\ell (t) $ proportional to $t^\beta/\log^{2\beta}t$. Although on a logarithmic scale the additional factor is smaller, we would like to do better.

The bound can be improved by considering a source that is somewhat smaller --- by a factor of two is sufficient -- than $\LT_n$ which increases the probability that it is filled, but means that more extra time, $T_I-T_S$, is needed to produce a mean number of jumps to the furthest bin of at least  $W$.   Using that $U$  and $\gamma \TT$ vary slowly with scale, one can expand $\Lambda(r,t)$ around $r=\LT_n$ and analyze the changes on the bounds at the next scale.   This improves the bounds to $\gamma \sim 1/(\sqrt{\beta}\TT)$ and $U \sim \beta^{5/2}$ with $\beta=1/(\mu-1)$.   The resulting lower bound on $\ell(t)$ is 
\be
\ell(t) > \tilde{A_\mu} \beta^{-\frac{5\beta}{2}} t^\beta
\ee
with the coefficient $\tilde{A_\mu} \sim 4^{\beta^2}$ that from the deterministic iteration which is the same, up to an order unity pre factor, as the upper bound.  
It is not clear where between the lower and upper bounds on the coefficient will be the typical behavior, nor how broad the fluctuations will be --- even  on a log scale. 

The behavior as $\mu \nearrow 2$ we have not analyzed explicitly,  instead focussing on the rapidly growing regime for $\mu \searrow 1$, but the bounds will be of similar form although more care is needed to get upper and lower bounds reasonably close to one another due to the important jumps being only a modest fraction of the size of the already occupied region.

For the marginal case, $\mu=1$,  one can find an upper bound on the time at which a the region out to $r$ is likely to be fully filled by similar methods to that for the intermediate-range power-law regime. This yields a bound $\LL^{\Phi=1}(t)$ of the same form as that above for partial filling ($\Phi<1$),, except with the coefficient of the $\log\log t$ term in  \eqref{low-bnd-marginal} equal to 6 instead of 2.   The convergence of the probability of being fully filled  is, however, much slower for this bound on compete filling than for the bound on being  mostly filled.  While the latter converges for $t>\TG^{\Phi}(r)\sim \TT(r)\sqrt{\log\TT(r)}\log\log\TT(r)$ with a rate of order unity --- dominated by the small scales --- the former converges as the time increases above $\TG^{\Phi=1}(r)\sim \TT(r) \log^{\frac{5}{2}}\TT(r)$ on a time scale, $1/\gamma$, of order $\TT(r) \sqrt{\log\TT(r)}$ ---  faster  than $\TG^{\Phi=1}(r)$ but not much so.   

Note that the convergence of the probability for being mostly filled to more than a fixed filling fraction, $\Phi$, is a hybrid property: the probability of a fixed site being filled by time $t$, $q(r,t)$, is bounded below by (roughly) the product of $\Phi$ and the probability that the region out to $r$ is filled to above $\Phi$.  To get the convergence of this to unity, $\Phi$ needs to be adjusted and the thus-far ignored $\log(1-\Phi)$ factors kept track of. This also necessitates treating intermediate scales differently as the number of bins that do not need to be filled, $(1-f_n)K_n$
is not large.   However a different approach would provide a better bound: focussing on a specific site being filled with high probability can be done by a method more analogous to the funnel picture of the main text:  for the site to be filled, it needs to be in a small-scale  bin that is mostly but not-necessarily fully filled with high probability, which needs itself to be in a larger bin similarly, etc. But these can be filled from source regions that are not fully filled: being partly filled with high enough probability is sufficient.  We have not carried out such analysis in detail in part because the actual mechanism by which sites that  are empty for anomalously long will be filled is more complicated as it will involve filling from nearby regions on a hierarchy of scales that were filled at more typical times.    

The analyses here can be immediately extended to give lower bounds on the average density profile at long distances: these will be of the same form as the lower bounds, thus demonstrating that the predicted $\frac{\ell(t)}{r}^{-(1+\mu)}$ is essentially correct. 

\subsubsection{Comparisons with results  of Chatterjee and Dey}

As noted in the main text, when this work was essentially complete, a
preprint by Chatterjee and Dey (CD) appeared which derives and proves
some results closely related to ours in the context of long-range
first passage percolation which is essentially equivalent to the
lattice dispersal model, with the jump kernel $G(r) \sim r^{-\alpha}$
equivalent to the $1/r^{-(d+\mu)}$ that we
  use~\cite{chatterjee2013multiple}.  While some of the quantities CD
  focus on are different, the leading asymptotic scaling behaviors
  they obtain are essentially the same, and their proofs apply in all
  dimensions.  Our $q(r,t)$ corresponds to the probability that the
  first passage time $T^F(r)$ is less than $t$ and their diameter,
  $D(t)$ --- the maximum distance between any pair of occupied points
  at time $t$ --- is, with high-probability that decays as a power of
  $ \ell/D$ --- not many times $\ell(t)$, as we both obtain.

  However CD's results are sub-optimal. In particular, for the
  coefficient, $C$, of $\log\ell(t)/\log^2 t$ in the marginal case,
  they only obtain upper and lower bounds instead of our exact result
  $C=1/4d\log 2$. Indeed, in one dimension we obtain upper and lower
  bounds on the errors: \be 1-c_-\frac{\log\log t}{\log t} <
  \frac{4\log2\log\ell(t)}{\log^2 t} < 1-\frac{\log\log t}{\log t} \ee
  with high probability --- in senses that can be made precise from
  our analysis --- with the coefficient $c_-$ either 2 or 6 depending
  on the definition of $\ell(t)$ used.  In the intermediate range
  power-law growth regime, CD's theorems do not appear to exclude
  $\log t$ pre-factors in $\ell(t)$, although their analysis might
  well do so.  But the main difference is our analysis of the whole
  crossover regime for $\mu$ near $d$, including the divergences and
  vanishings of coefficients of the asymptotic forms, which they do
  not consider.  These are crucial for comparisons with simulations
  because of the very long length scales of the crossovers.

To turn our upper and lower bounds into formal proofs in one dimension
requires primarily filling in some details associated with the small
scale regime. For higher dimensions some additional work is needed,
although the strategies should work without major modifications.




\section{Supplementary Figures}
\label{sec:suppl-figur}
\begin{figure}[ht!]
\center{\includegraphics[width=.9\columnwidth]{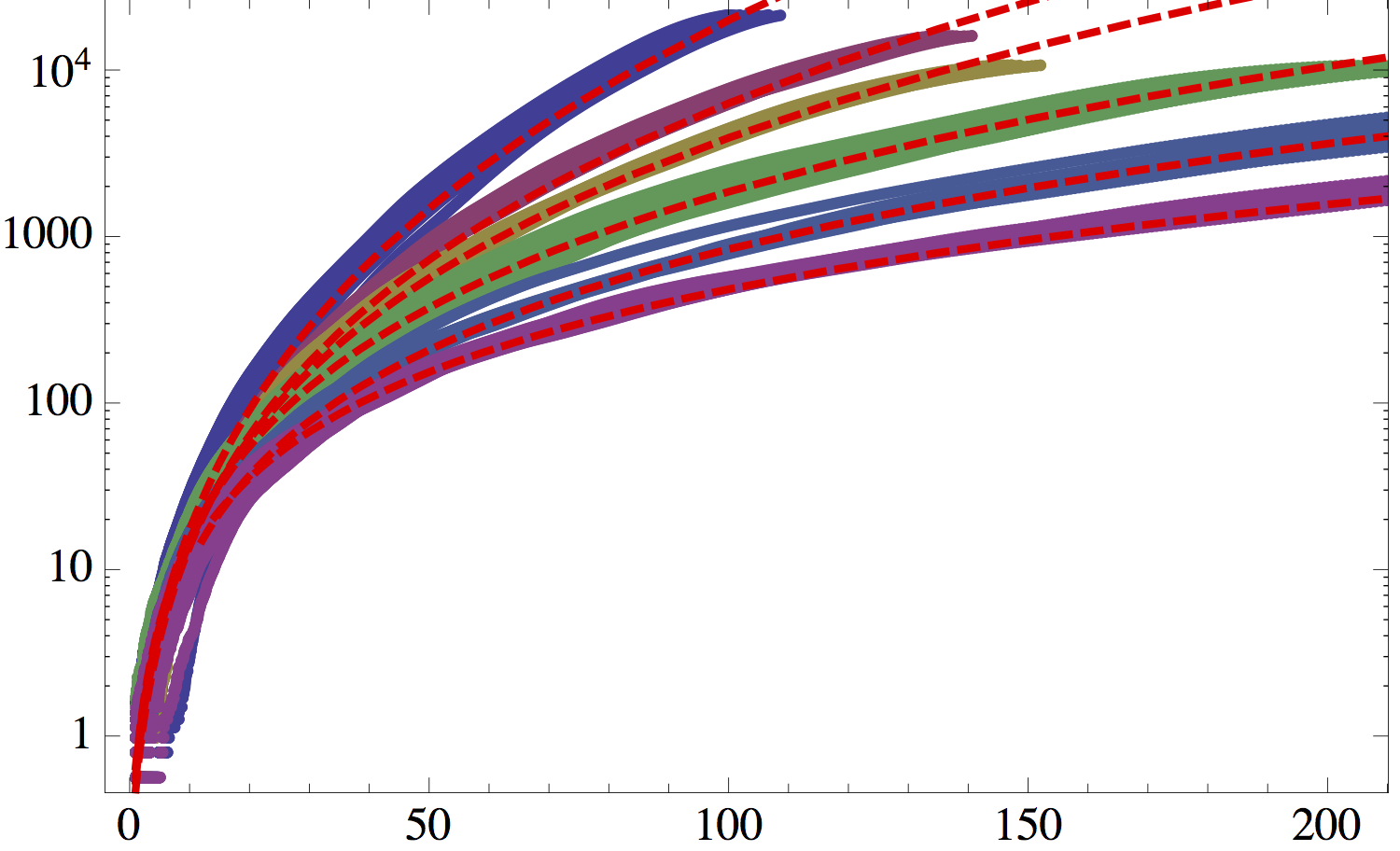}}
\caption{Summary of the spreading dynamics in two spatial dimensions.
  The effective radius $\ell(t)$ of the region occupied by the mutant
  population is plotted as a function of time $t$, for various
  long-range jump kernels. Each colored cloud represents data obtained
  from $10$ runs for a given jump kernel with tail exponent $\mu$ as
  indicated. Red dashed lines represent predictions, obtained from
  equation (8) (main Text) with fitted cross-over scales.  To speed up
  the simulations, we set $\tilde \epsilon=1$ (always jump) and $v=0$
  (no wavelike spread of clusters). The jump exponents are:
  $\mu\in\{1.6, 1.8, 1.9, 2.1, 2.3,
  2.5\}$.  \label{fig:two-dim-spreading}}
\end{figure}

\begin{figure}[ht!]
  \center{\includegraphics[width=.9\columnwidth]{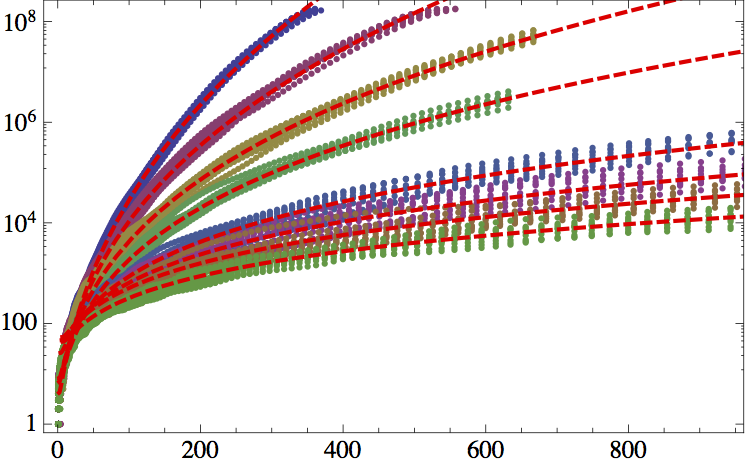}} 
  \caption{Summary of the spreading dynamics in one spatial dimensions
    with  short-range as well as long-range dispersal. For these simulations each cluster
    expands at a linear speed even in the absence of long-range
    jumps. In each time step, a long-range jump is performed only with
    probability $\tilde \epsilon=0.1$. For the short-range part, a pair of
    \emph{neighboring} sites is chosen at random. If this pair happens
    to fall on a boundary of a mutant cluster, i.e. the identity of
    both sites is mixed, then the wildtype site is switched to a
    mutant site. This leads to expansion of mutant clusters at average
    speed of $v_0=2$ sites per generation. The jump exponents are:
    $\mu\in\{0.6, 0.7, 0.8, 0.9, 1.0, 1.1, 1.2, 1.3, 1.4\}$; 
    The theoretical predictions indicated by the red dashed lines fit the data after choosing
    appropriate crossover time and length
    scales between linear growth and super-linear growth. \label{fig:v1-spreading-app}}
\end{figure}

\begin{figure} [ht!]
  \includegraphics[width=\columnwidth]{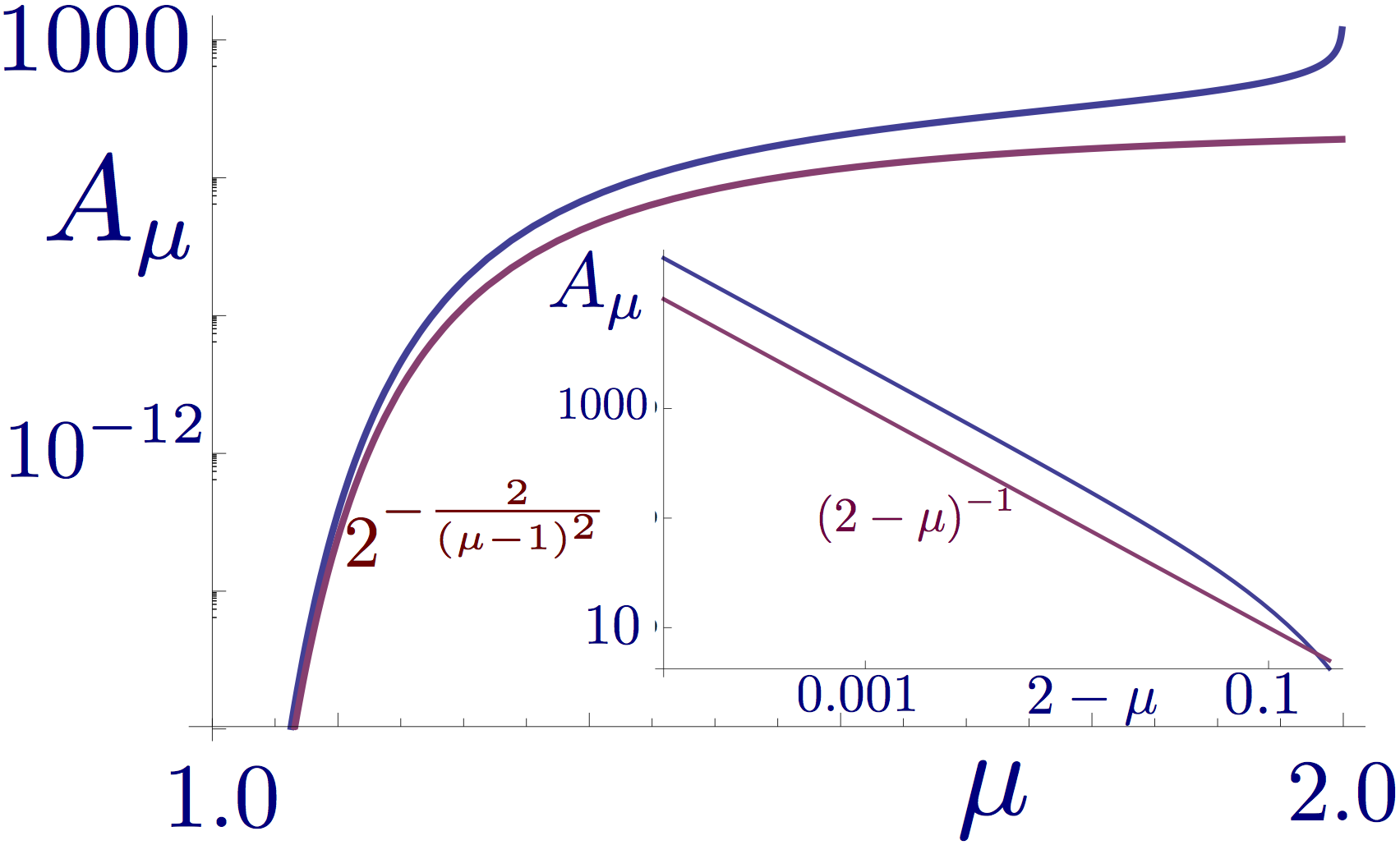}
  \caption{The pre-factor $A_\mu$ of the predicted power law growth in
    Eq.~(3) (main text) in one dimension: $\ell(t)\approx A_\mu
    t^\beta$ with $\beta=1/(\mu-1)$ for $1<\mu<2$. The blue curve is
    obtained numerically from solving equation (1) (main text) with
    the power-law ansatz; the red-curve represents an analytic
    approximation derived in ``Crossovers and Beyond Asymptopia''
    (main text). Notice the sharp (non-analytic) drop of the
    pre-factor as $\mu$ approaches $1$. The reason is very slow
    cross-over to the power law from an intermediate asymptotic regime
    controlled by the dynamics of the marginal case. As $\mu$
    approaches 2, the pre-factor diverges as $A_\mu\sim (2-\mu)^{-1}$
    indicative of another slow crossover at
    $\mu=2$, see Fig.~\ref{fig:second-marginal-point}. \label{fig:prefactors}}
\end{figure}

\begin{figure} [ht!]
  \includegraphics[width=\columnwidth]{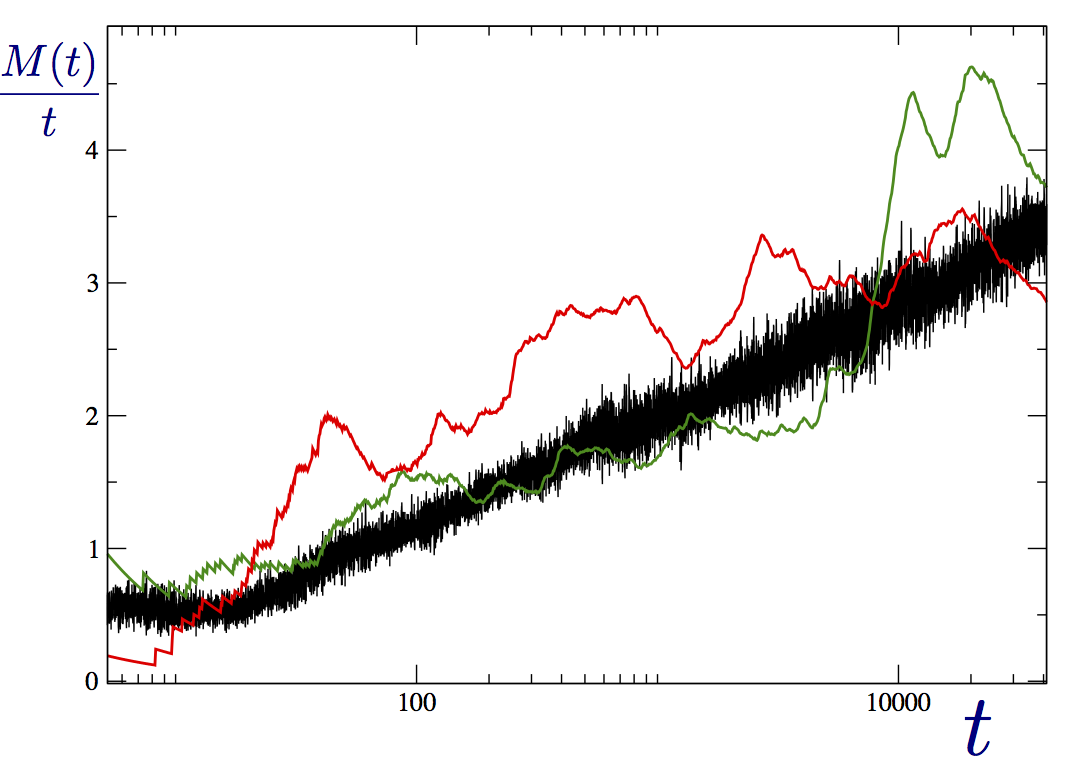}
  \caption{Dynamics of growth in one dimension at the marginal point
    between superlinear and linear growth, $\mu=2$. The number of
    mutant sites, $M(t)$, scaled by time, $M(t)/t$, is plotted
    as a function of time, averaged over 10 realizations (black) and
    for two individual realizations (red and green). While the
    averaged data suggests $M(t)\sim t \ln(t)$, the individual
    realizations indicate strong fluctuations caused by occasional
    rare jumps, which are of order $t$.  These ``leaps
    forward''~\cite{mollison1972rate} are driving the logarithmic
    increase of the spreading
    velocity. \label{fig:second-marginal-point}}
\end{figure}


\end{document}